\documentclass[sigconf]{acmart}

\AtBeginDocument{%
  \providecommand\BibTeX{{%
    \normalfont B\kern-0.5em{\scshape i\kern-0.25em b}\kern-0.8em\TeX}}}

\usepackage{mixedgcImp}
\begin{document}
\fancyhead{}

\title{Missing Value Imputation for Mixed Data via Gaussian Copula}

\author{Yuxuan Zhao, Madeleine Udell}
\affiliation{\institution{Cornell University}}
\email{{yz2295,udell}@cornell.edu}

\begin{abstract}
  Missing data imputation forms the first critical step of many data analysis pipelines.
	The challenge is greatest for mixed data sets, including real, Boolean, and ordinal data,
	where standard techniques for imputation fail basic sanity checks: for example, the imputed values may not follow the same distributions as the data.
	This paper proposes a new semiparametric algorithm to impute missing values,
	with no tuning parameters.
	The algorithm models mixed data as a Gaussian copula.
	This model can fit arbitrary marginals for continuous variables and can handle ordinal variables with many levels, including Boolean variables as a special case.
	We develop an efficient approximate EM algorithm to estimate
	copula parameters from incomplete mixed data.
	The resulting model reveals the statistical associations among variables.
	Experimental results on several synthetic and real datasets show
	the superiority of our proposed algorithm to state-of-the-art imputation algorithms for mixed data.
\end{abstract}

\ccsdesc[500]{Mathematics of computing~Expectation maximization}
\ccsdesc[500]{Mathematics of computing~Maximum likelihood estimation}
\ccsdesc[500]{Computing methodologies~Learning latent representations}

\keywords{mixed data, ordinal data, Gaussian copula, missing values, imputation}

\maketitle
\section{Introduction}

	Mixed data sets --- those that include real, Boolean, and ordinal data --- are a fixture of modern data analysis.
	Ordinal data is particularly common in survey datasets.
	For example, Netflix users rate movies on a scale of $1$-$5$.
	Social surveys may roughly bin respondents' income or level of education as an ordinal variable,
	and ordinal Likert scales measure how strongly a respondent agrees with certain stated opinions.
	Binary variables may be considered a special case of an ordinal with two levels.
	Health data often contains ordinals that result from patient surveys or from coarse binning of
	continuous data into, e.g., cancer stages 0--IV or overweight vs obese patients.
	In all of these settings, missing data is endemic due to nonresponse
	and usually represents a large proportion of the dataset.
    Missing value imputation generally precedes other analysis,
	since most machine learning algorithms require complete observations.
	Imputation quality can strongly influence subsequent analysis.
	
	To exploit the information in mixed data,
    imputation must account for the interaction between continuous and ordinal variables. 
    Unfortunately, 
    the joint distribution of mixed data can be complex.
    Existing parametric models are either too restrictive \cite{little2019statistical} or require priori knowledge of the data distribution \cite{van1999flexible}.
    Nonparametric methods, such as MissForest \cite{stekhoven2011missforest}, based on random forests, 
    and imputeFAMD \cite{audigier2016principal}, based on principal components analysis, tend to perform better.
    However, these two methods treat ordinal data as categorical, losing valuable information about the order.
    Further, they can only afford a limited number of categories.
	
	It is tempting, but dangerous, to treat ordinal data with many levels as continuous.
	For example, the ordinal variable ``Weeks Worked Last Year'' from the General Social Survey dataset takes $48$ levels, but $74\%$ of the population worked either 0 or 52 weeks.
    Imputation that treats this variable as continuous (e.g., imputing with the mean) works terribly!
	As another example, consider using low rank matrix completion \cite{candes2009exact,recht2010guaranteed,keshavan2010matrix,mazumder2010spectral} to impute missing entries in a movie rating datasets
	using a quadratic loss.
	This loss implicitly treats ratings encoded as $1$--$5$ as numerical values, so the difference between ratings 3 and 4 is the same as that between ratings 4 and 5. Is this true? How could we tell?
    
\begin{figure}[h]
		\centering
        \includegraphics[width=\linewidth]{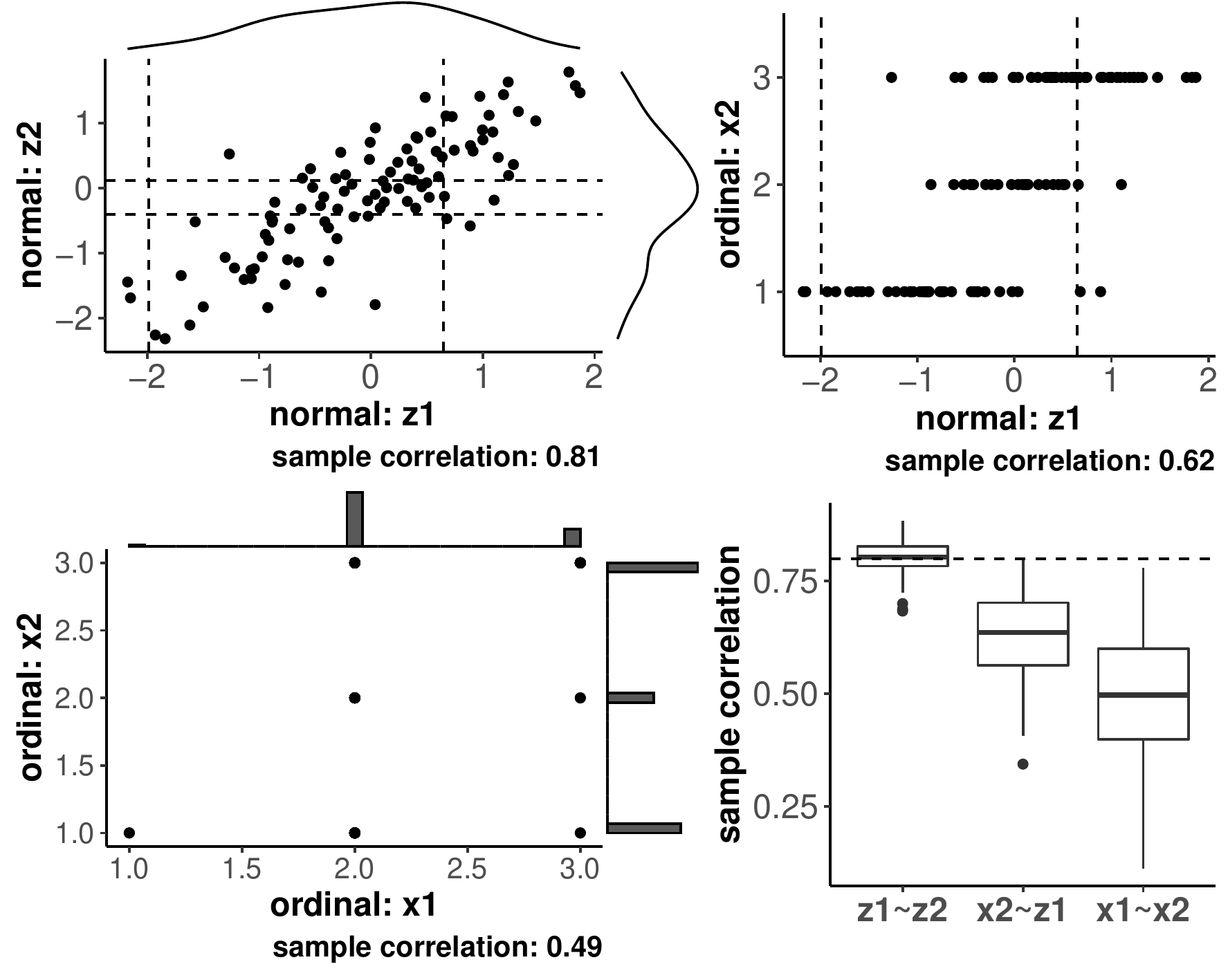}
		\caption{Draw $(z_1,z_2)$ from a binormal with correlation $0.8$. Discretize $z_1$ to $x_1$, $z_2$ to $x_2$ on random cutoffs. Top two and bottom left panels plot one repetition. Dashed lines mark the cutoffs. Bottom right panel plots the sample correlation over $100$ repetitions. Dashed line marks the truth.}
		\label{fig:intro}
	\end{figure}

    A more sensible (and powerful) model treats ordinal data as generated by
    thresholding continuous data, as in \cite{rennie2005loss,rennie2005fast}.
    Figure \ref{fig:intro} illustrates how correlations can by garbled by treating such data as continuous.
Our work builds on this intuition to model mixed data through the Gaussian copula model \cite{hoff2007extending,liu2009nonparanormal,fan2017high,feng2019high},
which assumes the observed vector is generated by transforming each marginal of a latent normal vector.
Under this model,
we associate each variable (both ordinal and continuous) with a latent normal variable.
Each ordinal level corresponds to an interval of values of the corresponding latent normal variable.

We propose an efficient EM algorithm to 
	estimate a Gaussian copula model with incomplete mixed data
	and show how to use this model to impute missing values. 
    Our method outperforms many state-of-the-art imputation algorithms for various real datasets including social survey data (whose columns have a varying number of ordinal levels), movie rating data (high missing ratio), music tagging data (binary data), etc.
	The proposed method has several advantages: 
	the method has no hyper-parameters to tune
	and is invariant to coordinate-wise monotonic transformations in the data.
	Moreover, 	
	the fitted copula model is interpretable and can reveal statistical associations among variables, which is useful for social science applications.	
	To our best knowledge, 
    our proposed algorithm is the first frequentist approach to fit the Gaussian copula model with incomplete mixed data.
    Moreover, it is much faster than the existing Bayesian MCMC algorithm for the same model \cite{hoff2007extending}; given the same time budget, our method produces substantially more accurate estimates.

\subsection{RELATED WORK}
\paragraph{Gaussian Copula for Mixed Data.}
Modeling mixed data with the Gaussian copula model has been studied using both frequentist approaches \cite{fan2017high, feng2019high} and Bayesian approaches
\cite{hoff2007extending, murray2013bayesian, cui2019novel}.
In \cite{murray2013bayesian,cui2019novel}, 
the authors further assume the latent normal vector is generated from a factor model.
When all variables are ordinal, 
the Gaussian copula model is equivalent to the probit graphical model \cite{guo2015graphical}.
However, all these previous work focuses on model estimation and theoretical properties of the estimators, 
and has overlooked the potential of these models for missing value imputation.

In fact, the frequentist parameter estimation methods proposed \cite{fan2017high,feng2019high,guo2015graphical}
assume complete data;
so these methods cannot perform imputation.
Among Bayesian approaches,
MCMC algorithms \cite{hoff2007extending, murray2013bayesian, cui2019novel} can fit the copula model with incomplete data and impute missing values.
However, to use these models, one must select the number of factors for the models in \cite{ murray2013bayesian, cui2019novel}.
The sensitivity of these models to this parameter makes it a poor choice in practice for missing value imputation.


The implementation of \cite{hoff2007extending} is still the best
method available to fit a Gaussian copula model for incomplete mixed data.
\citet{hollenbach2018multiple} provides an important case study of this method for use in multiple imputation with an application to sociological data analysis. 
However, the method is slow and sensitive: the burn-in and sampling period must be carefully chosen for MCMC to converge, 
and many iterations are often required, so the method does not scale to even moderate size data, which limits its use in practice.
Our model matches that of \cite{hoff2007extending}, but our EM algorithm runs substantially faster.

\paragraph{Low rank matrix completion.}
The generalized low rank models framework \cite{udell2016generalized} handles missing values imputation for mixed data using a low rank model with appropriately chosen loss functions
to ensure proper treatment of each data type.
However, choosing the right loss functions for mixed data is challenging.
A few papers share our motivation:
for example, early papers by Rennie and Srebro \cite{rennie2005loss,rennie2005fast} proposed a thresholding model
to generate ordinals from real low rank matrices.
\citet{ganti2015matrix} estimate monotonic transformations of a latent low rank matrix,
but the method performs poorly in practice.
\citet{anderson2018xpca} posits that the mixed data are generated by marginally transforming the columns of the sum of a low rank matrix and isotropic Gaussian noise. 
While their marginal transformation coincides with the Gaussian copula model,
their setup greatly differs in that it cannot identify the correlations between variables. 

While low rank matrix completion methods scale well to large datasets, 
the low rank assumption is too weak to generalize well on long skinny datasets.
Hence low rank methods tend to work well on ``square-ish'' datasets ($n \sim p$) \cite{udell2019why},
while the copula methods proposed here work better on long, skinny datasets.

	\section{Notation}
	\label{sec:notation}
	Define $[p]=\{1,\ldots,p\}$ for $p\in \mathbb{Z}$.
	Let $\bx=(x_1,\ldots,x_p)\in \mathbb{R}^{p}$ be a random vector.
	We use $\bx_{I}$ to denote the subvector of $\bx$ with entries in subset $I\subset [p]$.
	Let $\indMis, \indexCobs, \indexDobs \subset [p]$ denote missing,
	observed continuous, and observed discrete (or ordinal) dimensions, respectively.
    The observed dimensions are $\indexO=\indexC\cup \indexD$, so $\bx=(\bx_{\indexCobs},\bx_{\indexDobs}, \xmis)=(\xobs,\xmis)$.

	Let $\bX\in \mathbb{R}^{n\times p}$ be a matrix whose rows correspond to observations
	and columns to variables.
	We refer to the $i$-th row, $j$-th column, and $(i,j)$-th element as $\bx^i, \bX_j$ and $x_j^i$, respectively.

	We say random variables $x= y$ and random vectors $\bx=\mathbf{y}$ if
	their cumulative distribution functions (CDF) match.
	The elliptope 
	$\mathcal E = \{Z \succeq 0: \text{diag}(Z) = 1\}$
	is the set of correlation matrices.

	\section{Gaussian Copula}
	The Gaussian copula models complex multivariate distributions through
	transformations of a latent Gaussian vector.
	We call a random variable $x\in\mathbb{R}$ continuous when it is supported on an interval. We can match the marginals of any continuous random vector $\bx$
	by applying a strictly monotone function to a random vector $\bz$
	with standard normal marginals.
	Further, the required function is unique, as stated in \cref{lemma:marginalTransform}.

		\begin{lemma}
		Suppose $\bx\in \mathbb{R}^p$ is a continuous random vector with CDF $F_j$ for each coordinate $j\in[p]$,
		and $\bz\in \mathbb{R}^p$ is a random vector with standard normal marginals.
		Then there exists a unique
		elementwise strictly monotone function
		$\bigf(\bz) := (f_1(z_1),\ldots,f_p(z_p))$
		such that
		\begin{equation}
	x_j = f_j(z_j) \quad \mbox{and} \quad f_j=F_j^{-1}\circ \Phi, \quad j \in[p],
	\label{Eq:functionf}
		\end{equation}
		where $\Phi$ is the standard normal CDF.
		\label{lemma:marginalTransform}
	\end{lemma}

All proofs appear in the supplementary materials.
Notice the functions $\{f_j\}_{j=1}^p$ in \cref{Eq:functionf} are strictly monotone,
so their inverses exist.
Define $\bigf^{-1}=(f_1^{-1},\ldots,f_p^{-1})$.
Then $\bz =\bigf^{-1}(\bx)$ has standard normal marginals,
but the joint distribution of $\bz$ is not uniquely determined.
The Gaussian copula model (or equivalently nonparanormal distribution \cite{liu2009nonparanormal}) further assumes $\bz$ is jointly normal.

	\begin{definition}
		We say a continuous random vector $\bx\in \mathbb{R}^{p}$ follows
		the Gaussian copula $\bx\sim \textup{GC}(\Sigma,\bigf)$
		with parameters $\Sigma$ and $\bigf$
		if there exists a correlation matrix $\Sigma$
		and elementwise strictly monotone function
		$\bigf: \reals^p \to \reals^p$
		such that $\bigf(\bz)=\bx$ for $\bz \sim \mathcal{N}_p(\mathbf{0},\Sigma)$.
	\end{definition}

This model is semiparametric:  it comprises nonparametric functions $\bigf$ and parametric copula correlation matrix $\Sigma$.  The monotone $\bigf$ establishes the mapping between observed $\bx$ and latent normal $\bz$, while $\Sigma$ fully specifies the distribution of $\bz$.
Further, the correlation $\Sigma$ is invariant to
elementwise strictly monotone transformation of $\bx$.
Concretely, if $\bx\sim \textup{GC}(\Sigma,\bigf)$ and $\by= \mathbf{g}(\bx)$
where $\mathbf{g}$ is elementwise strictly monotone,
then $\by \sim  \mbox{GC}(\Sigma,\bigf\circ \mathbf{g}^{-1})$.
Thus the Gaussian copula separates the multivariate interaction $\Sigma$
from the marginal distribution $\bigf$.

\begin{figure}[h]
	\centering{
	\begin{tikzpicture}[yscale = 0.75, xscale =0.75]
	\begin{axis}[
	ylabel = \textbf{ordinal x value},
	xlabel = \textbf{normal z value},
	xtick = {-3,-1,0,1,3},
	xticklabels = {$\mathbf{-\infty}$, $\mathbf{-1}$, $\mathbf{0}$, $\mathbf{1}$, $\mathbf{-\infty}$},
	ytick = {1,2,3},
	yticklabels = {$\mathbf{1}$, $\mathbf{2}$, $\mathbf{3}$}
	]
	\addplot+[jump mark right, black, line width=0.5mm]  coordinates {
		(-1,1) (1,2)
	};
	\addplot [line width=0.5mm, black, domain=1:3, samples=100] {3};
	\addplot [line width=0.5mm, black, domain= -3:-1, samples=100] {1};
	\addplot [line width=0.5mm, black, dashed] coordinates {
		(-1,1) (-1,2)
	};
	\addplot [line width=0.5mm, black, dashed] coordinates {
		(1,2) (1,3)
	};
	\addplot [line width=0.5mm, black] coordinates {
		(-1,2) (1,2)
	};
	\end{axis}
	\end{tikzpicture}
}
	\caption{Cutoff function $f(\cdot)$ with cutoffs $\{-1,1\}$ maps continuous $z$ to ordinal $x\in \{1,2,3\}$.}
	\label{fig:cutoff}
\end{figure}
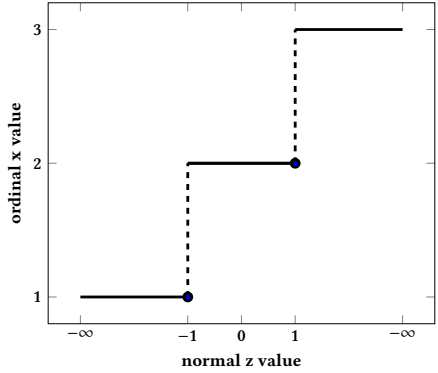

When $f_j$ is strictly monotone, $x_j$ must be continuous.
On the other hand, when $f_j$ is monotone but not strictly monotone,
$x_j$ takes discrete values in the range of $f_j$ and can model ordinals.
Thus for ordinals, $f_j$ will not be invertible.
For convenience, we define a set-valued inverse $f_j^{-1}(x_j):=\{z_j:f_j(z_j)=x_j\}$.
When the ordinal $x_j$ has range $[k]$, \cref{lemma:cutoff} states that
the only monotone function $f_j$ mapping continuous $z_j$ to $x_j$ is a cutoff function,
defined for some parameter $S \subset \reals$ as
\[
\cutoff(z; \bS):=1+\sum_{s\in \bS}\mathds{1}(z > s) \mbox{ for }z\in\mathbb{R}.
\]
\begin{lemma}
		Suppose $x\in \mathbb{R}$ is an ordinal random variable with range $[k]$ and probability mass function $\{p_l\}_{l=1}^k$
		and $z\in \mathbb{R}$ is a continuous random variable with CDF $F_z$.
		Then $f=\cutoff(z; \bS)$ is the unique monotone function $f$ that satisfies $x=f(z)$,
		where $\bS=\{s_l = F_z^{-1}\left( \sum_{t=1}^lp_t\right): l \in [k-1]\}$.
		\label{lemma:cutoff}
	\end{lemma}
For example, in recommendation system we can think of the discrete ratings as obtained by rounding some ideal real valued score matrix. The rounding procedure amounts to apply a cutoff function. See Figure \ref{fig:cutoff} for an example of cutoff function.

\begin{figure}[h]
		\centering
		\includegraphics[width=0.8\linewidth]{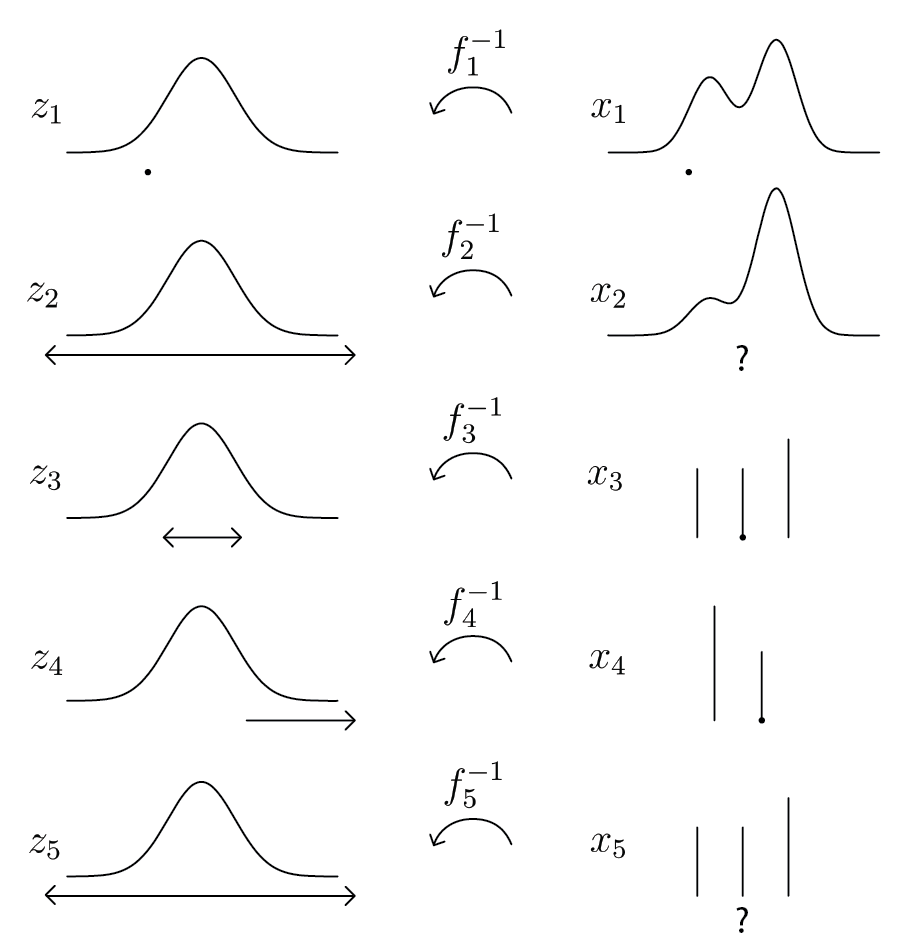}
		\caption{An example of $5$ dimensional Gaussian copula model.
		For observed continuous $x_1$, the corresponding $z_1$ takes a fixed value. 
		For observed ordinal $x_3$ and $x_4$, the corresponding $z_3$ and $z_4$ take values from an interval. 
		For missing continuous $x_2$ and missing ordinal $x_5$, the corresponding $z_2$ and $z_5$ can take any value.}
		\label{fig:copula}
	\end{figure}

To extend the Gaussian copula to mixed data, we simply specify that $f_j$ is strictly monotone for $j\in\indexC$ and that $f_j$ is a cutoff function for $j\in\indexD$.
As before, the correlation $\Sigma$ remains invariant to elementwise strictly monotone transformations.
The main difference is that while $f_j^{-1}(x_j)$ is a single number when $j \in \mathcal C$ is continuous,
it is an interval when $j \in \mathcal D$ is discrete. See Figure \ref{fig:copula} for illustration.

	\section{Imputation}
\label{sec:imp}
So far we have introduced a very flexible model for mixed data.
Our interest is to investigate missing value imputation under this model. 
Concretely,
suppose the data matrix $\bX$ has rows $\bx^1,\ldots,\bx^n\overset{i.i.d.}\sim \textup{GC}(\Sigma,\bigf)$ and $\bx^i=(\xic, \xid, \bx^i_{\indiMis})=(\xio,\bx^i_{\indiMis})$ for $i\in[n]$. 
we first estimate $\hat \Sigma$ and $\hat \bigf$ using observation $\{\xio\}_{i=1}^n$ 
and then impute missing values  $\{\xim\}_{i=1}^n$ using $\hat \Sigma$, $\hat \bigf$ and observation $\{\xio\}_{i=1}^n$.

In this section we first show how to impute the missing values with given estimates $\hat\bigf$ and $\hat\Sigma$.
The estimation for $\bigf$ appears in Section \ref{section:estF}, 
and the estimation for $\Sigma$ appears in Section \ref{section:estCov}.
The missing completely at random (MCAR) assumption is needed to consistently estimate $\bigf$. If the true $\bigf$ is known, the missing at random (MAR) assumption suffices to consistently estimate $\Sigma$. 
We discuss this issue further later in the paper.

For the latent normal vector $\bz^i$ satisfying $\bx^i=\bigf(\bz^i)$,
$\bz^i$ follows truncated normal distribution.
Define $\bigf_I=(f_j)_{j\in I}$ for $I\subset[p]$ and $f^{-1}_j(x^i_j)=\mathbb{R}$ for $j \in \indiMis$. 
In observed continuous dimensions $\indexC_i$, $\zic$ reduces to the point $\bigf_{\indexC_i}^{-1}(\xic)$.
In observed ordinal dimensions $\indexD_i$, $\zid$ lies in the Cartesian product of intervals  $\bigf_{\indexD_i}^{-1}(\xid)$.
There is no constraint in missing dimension $\indexM_i$.
It is natural to impute $\xim$ by mapping the conditional mean of $\zimis$ through the marginals $\bigf_{\indexM_i}$, summarized in \cref{alg:impute}.

	\begin{algorithm}
		\caption{Imputation via Gaussian Copula}
		\begin{spacing}{0.7}
		\begin{algorithmic}
			\STATE \textbf{Input:} observation $\{\xio\}_{i=1}^n$, 
			parameters estimate $\hat\bigf^{-1}$ and $\hat \Sigma$.
			\begin{enumerate}[itemsep=0mm]
				\item Compute constraints $\ziobs \in \hat \bigf_{\indiObs}^{-1}(\bx^i_{\indiObs}), i\in[n]$.
				\item For $i=1,\ldots,n$,
				\begin{itemize}[itemsep=0mm,topsep=0pt]
				\item Impute $\hat\bz^i_{\indiMis}=\Erm[\zimis|\ziobs \in \hat \bigf_{\indiObs}^{-1}(\bx^i_{\indiObs}),\hat \Sigma]$.
				\item Impute $\hat\bx^i_{\indiMis}=\hat\bigf_{\indiMis}(\hat\bz^i_{\indiMis})$.
				\end{itemize}
			\end{enumerate}
			\textbf{Output:} $\hat\bx^i_{\indiMis}$ for $i\in[n]$. 
		\end{algorithmic}
	\end{spacing}
	\label{alg:impute}
	\end{algorithm}


	While most applications require just a single imputation,
	multiple imputations are useful to describe the uncertainty due to imputation. 
	Our method also supports multiple imputation: in step (2) of Algorithm \ref{alg:impute}, replace the conditional mean imputation with conditional sampling and then impute $\hat\bx ^i_{\indiMis}$ for each sample. The conditional sampling consists of two steps: 
	(1) sample the truncated normal $\ziobs$ conditional on $\xio$ and $\hat \Sigma$; 
	(2) sample the normal $\zimis$ conditional on $\ziobs$ and $\hat \Sigma$. 
	Efficient sampling methods have been proposed \citep{pakman2014exact} for multivariate truncated normal distribution.

	\section{Monotonic Function Estimation}
	\label{section:estF}
	To map between $\bx$ and $\bz$, we require both $\bigf^{-1}$ and $\bigf$. It is easier to directly estimate $\bigf^{-1}$.
	For $j \in \indexC$, we have $f_j^{-1}=\Phi^{-1}\circ F_j$, as shown in \cref{Eq:functionf}.
	While the true CDF $F_j$ is usually unavailable, it is natural to estimate it by the
	empirical CDF of $\bX_j$ on the observed entries, denoted as $\hat  F_{j}$.
	We use the following estimator:
	\begin{equation}
	\hat f_j^{-1}(x^i_j)=\Phi^{-1}\left(\frac{n}{n+1}\hat F_{j}(x^i_j)\right).
	\label{Eq:estFcontinuous}
	\end{equation}
	The scale constant $n/(n+1)$ ensures the output is finite. MCAR assumption guarantees the observed entries of $\bX_j$ are from the distribution of $F_j$. Consider a case when MCAR is violated: an entry is observed if and only if it is smaller than a constant $c$, then the observed entries are actually from the distribution $\tilde F_j$:
	\[
	\tilde F_j(x_j)=\begin{cases} 
      F_j(x_j)/F_j(c), & \mbox{ when }x\leq c \\
      1, & \mbox{ when }x > c
   \end{cases}.
	\]
	Thus we assume MCAR in this section. 
	This assumption may be relaxed to MAR or even missing not at random by carefully modeling $F_j$ or the missing mechanism. 
	We leave that to our future work. \cref{Lemma:ContinuousConsistent} shows this estimator converges to $f_j^{-1}$ in sup norm on the observed domain.
	\begin{lemma}
		Suppose the continuous random variable $x \in \reals$ with CDF $F_x$ and
		normal random variable $z \in \reals$ satisfy $f(z){=}x$ for a strictly monotone $f$.
		Given $x^1,\ldots, x^n\overset{i.i.d.}\sim F_x$,
		$m=\min\limits_ix^i$, and $M=\max\limits_ix^i$,
		the inverse $\hat f^{-1}$ defined in \cref{Eq:estFcontinuous} satisfies
		\[
		\mathrm{P}\left( \sup_{m\leq x \leq M}|\hat f^{-1}(x) - f^{-1}(x)|> \epsilon\right) \leq  2e^{-{c_1n\epsilon^2}},
		\]
        for any $\epsilon$ in ${a_1}{n}^{-1}<\epsilon < b_1$,
		where $a_1, b_1, c_1>0$ are constants depending on $F_x(m)$ and $F_x(M)$.
		\label{Lemma:ContinuousConsistent}
	\end{lemma}

	For an ordinal variable $j \in \mathcal D$ with $k$ levels,
	$f_j(z_j)=\cutoff(z_j; \bS^{j})$.
	Since $\bS^{j}$ is determined by the probability mass function $\{p_l^j\}$ of $x_j$,
	we may estimate cutoffs $\hat \bS^{j}$ as a special case of \cref{Eq:estFcontinuous}
	by replacing $p^j_l$ with its sample mean:
	\begin{equation}
	\bS^{j} = \left\{\Phi^{-1}\left(\frac{\sum_{i=1}^{n_j}\mathds{1}(x^i_{j}\leq l)}{n_j+1}\right), \; l\in [k-1]\right\}.
		\label{Eq:estFordinal}
	\end{equation}
    Lemma \ref{Lemma:CutoffConsistent} shows that $\hat \bS^{j}$ consistently estimates $\bS^{j}$.
\begin{lemma}
		Suppose the ordinal random variable $x \in [k]$
		with probability mass function $\{p_l\}_{l=1}^k$
		and normal random variable $z \in \reals$ satisfy $f(z) = \cutoff(z; \bS) {=}x$.
	Given samples $x^1,\cdots, x^n \overset{i.i.d.}\sim \{p_l\}_{l=1}^k$,
	the cutoff estimate $\hat\bS$ from \cref{Eq:estFordinal} satisfies
	\[
\mathrm{P}\left( 	||\hat\bS - \bS||_1 > \epsilon\right)\leq   2^ke^{- c_2n\epsilon^2/(k-1)^2},
	\]
	 for any $\epsilon$ in $(k-1){a_2}{n}^{-1}<\epsilon < (k-1)b_2$,
	 where $a_2, b_2, c_2>0$ are constants depending on $\{p_1,p_k\}$.
	\label{Lemma:CutoffConsistent}
\end{lemma}

	\section{Copula Correlation Estimation}
	\label{section:estCov}
	We first consider maximum likelihood estimation (MLE) for $\Sigma$ with complete continuous observation, then generalize the estimation method to incomplete mixed observation. 

	\subsection{Complete Continuous Observations}
	We begin by considering continuous, fully observed data: $\indexD = \indexM = \emptyset$.
	The density of the observed variable $\bx$ is
	\begin{equation*}
		p(\bx;\Sigma,\bigf)\; d\bx 
		= \phi(\bz;\Sigma) d\bz,
	\end{equation*}
	where $\bz = \bigf^{-1}(\bx), d\bz=\left|\frac {\partial \bz}{\partial \bx}\right|d\bx$,
	$\phi(\cdot; \Sigma)$ is the PDF of the normal vector
	with mean $\bo$ and covariance $\Sigma$.
	The MLE of $\Sigma$ maximizes the likelihood function defined as: 
	\begin{align}
	        \ell(\Sigma;\bx^i)&=\frac{1}{n}\sum_{i=1}^n\log \phi(\bigf^{-1}(\bx^i);\Sigma)\nonumber\\
    &=c-\frac{1}{2}\log\det \Sigma -\frac{1}{2}\mbox{Tr}\left(\Sigma^{-1} \frac{1}{n}\sum_{i=1}^n\zi (\zi)^\intercal \right), \label{Eq:Contlikelihood}
	\end{align}
	 over $\Sigma \in \mathcal E$, where $\zi=\bigf^{-1}(\bx^i)$ and $c$ is a universal constant  
	 (We omit here and later the constant arising from $\left|\frac {\partial \bz}{\partial \bx}\right|$ after the log transformation).
	 Thus the MLE of $\Sigma$ is the sample covariance of $\bZ:=\bigf(\bX)=[f_1(\bX_1),\ldots,f_p(\bX_p)]$. 
	 When we substitute $\bigf$ by its empirical estimation in Eq. (\ref{Eq:estFcontinuous}),  the resulting covariance matrix $\tilde \Sigma$ of $\hat \bZ:=\hat\bigf(\bX)$ is still consistent and asymptotically normal under some regularity conditions \cite{tsukahara2005semiparametric},
	 which justifies the use of our estimator $\hat\bigf$.
	 To simplify notation, we assume $\bigf$ is known below. 
	 
	 For a Gaussian copula, notice $\Sigma$ is a correlation matrix, thus we update $\hat\Sigma = P_{\mathcal E} \tilde\Sigma$,
     where $P_{\mathcal E}$ scales its argument to output a correlation matrix:
     for $D = \text{diag}(\Sigma)$, $P_{\mathcal E}(\Sigma) = D^{-1/2} \Sigma D^{-1/2}$. The obtained $\hat \Sigma$ is still consistent and asymptotically normal.

\subsection{Incomplete Mixed Observations}
	\label{section:IncompleteMixed}
When some columns are ordinal and some data is missing, 
the Gaussian latent vector $\zi$ is no longer fully observed. 
We can compute the entries of $\zi$ 
corresponding to continuous data:  $\zic=\fic^{-1}(\xic)$.
However, for ordinal data, 
$\fid^{-1}(\xid)$ is a Cartesian product of intervals; we only know that $\zid \in \fid^{-1}(\xid)$.
The entries corresponding to missing observations, $\zimis$, are entirely unconstrained.
Hence the latent matrix $\hat \bZ$ is only incompletely observed, and it is no longer possibly to simply compute its covariance.

We propose an expectation maximization (EM) algorithm to estimate $\Sigma$ for incomplete mixed observation. Proceeding in an iterative fashion, we replace unknown $\zi (\zi)^\intercal$ with their expectation conditional on observations $\xio$ and an estimate $\hat \Sigma$ in the E-step, then in the M-step we update the estimate of $\Sigma$ as the conditional expectation of covaraince matrix:
	\begin{equation} \label{Eq:G-func}
	G(\hat\Sigma, \xio) = \frac 1 n \sum_{i=1}^n \Erm[\bz^i (\bz^i)^\intercal| \xio,\hat\Sigma].
	\end{equation}
Similar to the case of complete continuous data, we further scale the estimate to a correlation matrix. We first present the EM algorithm in Algorithm \ref{alg:EM}, then provide precise statements in Section \ref{section:em}. Computation details of Algorithm \ref{alg:EM} appear in Section \ref{sec:comp_em} and Section \ref{sec:comp_ordinal}.
 	\begin{algorithm}
 	\caption{EM algorithm for Gaussian Copula}
 	\begin{algorithmic}
 		\STATE \textbf{Input:} observed entries $\{\bx^i_{\indexO_i}\}_{i=1}^n$.
 		\STATE \textbf{Initialize:} $t=0$, $\Sigma^{(0)}$.
 		\STATE For $t = 0, 1, 2, \ldots$
 		\begin{enumerate}[itemsep=0mm]
 			\item E-step: Compute $G^{(t)} = G(\SigmaT, \bx^i_{\indexO_i})$. 
 			\item M-step: $\Sigma^{(t+1)}=G^{(t)}$.
 			\item Scale to correlation matrix: $\Sigma^{(t+1)}=P_{\mathcal E}(\Sigma^{(t+1)})$ 
 		\end{enumerate}
 		until convergence. \\
 		\textbf{Output:} $\hat{\Sigma}=\SigmaT$.
 	\end{algorithmic}
 	\label{alg:EM}
 \end{algorithm}

\subsection{EM algorithm}
\label{section:em}
We first write down the marginal density of observed values by integrating out the missing data. Since $\bx^i\sim \textup{GC}(\Sigma,\bigf)$, there exist latent $\bz^i$ satisfying $\bigf(\bz^i)=\bx^i$ and $\bz^i\sim \mathcal{N}_p(\bo,\Sigma)$. 
The likelihood of $\Sigma$  given observation $\xio$ is the integral
over the latent Gaussian vector $\ziobs$
that maps to $\xio$ under the marginal $\bigf_{\indexO_i}$.
Hence the observed log likelihood we seek to maximize is:
\begin{equation}
	 \ell_{\textup{obs}}(\Sigma; \xio)=\frac{1}{n}\sum_{i=1}^n\int_{\ziobs \in \bigf^{-1}_{\indexO_i}(\xio)}	\phi(\ziobs;\bo,\Sigma_{\indexO_i,\indexO_i})\; d\ziobs,
	 \label{Eq:likelihoodMixed}
\end{equation}
where $\Sigma_{\indexO_i,\indexO_i}$ denote the submatrix of $\Sigma$ with rows and columns in $\indexO_i$.
With known $\bigf$, MAR mechanism guarantees the maximizer of the likelihood in Eq. (\ref{Eq:likelihoodMixed}) shares the consistency and asymptotic normality of standard maximum likelihood estimate, according to the classical theory \cite[Chapter~6.2]{little2019statistical}.

However, the maximizer has no closed form expression. Even direct evaluation of $\ell_{\textup{obs}}(\Sigma;\bx^i_{\indiObs})$ is challenging since it involves multivariate Gaussian integrals in a truncated region and the observed locations $\indiObs$ varies for different observations $i$. Instead, the proposed EM algorithm is guaranteed to monotonically converge to a local maximizer according to classical EM theory  \cite[Chapter~3]{mclachlan2007algorithm}. 

	Now we derive the proposed EM algorithm in detail.
	Suppose we know the values of the unobserved $\bz^i$. Then the joint likelihood function is the same as in Eq. (\ref{Eq:Contlikelihood}).
    Since the values of $\bz^i$ are unknown, we treat $\zi$ as latent variables and $\xio$ as observed variables. Substituting the joint likelihood function by its expected value given observations $\bx^i_{\indObs}$ and an estimate $\hat\Sigma$:
		\begin{align*}
	&Q(\Sigma;\hat\Sigma, \xio):=\frac{1}{n}\sum_{i=1}^n\Erm[\ell(\Sigma;\bx^i_{\indiObs}, \bz^i)| \xio,\hat\Sigma] \\
	&= c- \frac{1}{2}\left( \log \det (\Sigma)+\mbox{Tr}\left(\Sigma^{-1} G(\hat\Sigma, \xio) \right) \right).\label{Eq:Qfunction}
	\end{align*}
EM theory \cite[Chapter~3]{mclachlan2007algorithm} guarantees the updated \\$\tilde\Sigma =\argmax_{\Sigma \in \mathcal E}Q(\Sigma;\hat\Sigma, \xio)$
	improves the likelihood with $\hat\Sigma$,
	\[
\ell_{\textup{obs}}(\tilde\Sigma;\bx^i_{\indiObs}) \geq \ell_{\textup{obs}}(\hat\Sigma;\bx^i_{\indiObs}),
	\]
and that by iterating this update, 
we produce a sequence $\{\SigmaT\}$ that converges monotonically to a local maximizer of $\ell_{\textup{obs}}(\Sigma;\bx^i_{\indiObs})$.
 At the $t$-th iteration, for the E step we compute $\Erm[\bz^i(\bz^i)^\intercal| \xio,\SigmaT]$ to express $Q(\Sigma; \Sigma^{(t)}, \xio)$ in terms of $\Sigma$. For the M step, we find $\Sigma^{(t+1)}=\argmax_\Sigma  Q(\Sigma ;\Sigma^{(t)}, \xio)$. In practice, we resort to an approximation, as in \cite{guo2015graphical}.
Notice that the unconstrained maximizer is $\tilde\Sigma=G(\SigmaT,\xio)$.
We update $\Sigma^{(t+1)} = P_{\mathcal E} \tilde\Sigma$.



\subsection{Conditional Expectation Computation}
\label{sec:comp_em}
Suppressing index $i$, we now show how to compute $\Erm[\bz\bz^\intercal| \xobs,\Sigma]$ in Eq. (\ref{Eq:G-func}). With $\bz_{\indexC}=\bigf_{\indexC}^{-1}(\bx_{\indexC})$, it suffices to compute the following terms:
		\begin{enumerate}[itemsep=0mm]
		\item the conditional mean and covariance of observed ordinal dimensions $\Erm[\bz_{\indexD}| \xobs,\Sigma], \Covrm[\bz_{\indexD}| \xobs,\Sigma]$.
		\item the conditional mean and covariance of missing dimensions  $\Erm[\bz_\indMis| \xobs,\Sigma], \Covrm[\bz_\indMis| \xobs,\Sigma]$.
		\item the conditional covariance between missing and observed ordinal dimensions $\Covrm[\bz_\indMis,\bz_\indexD| \xobs,\Sigma]$.
	\end{enumerate}
	We show that with the results from (1), we can compute (2) and (3). Computation for (1) is put in Sec \ref{sec:comp_ordinal}.
	
	Suppose we can know the ordinal values $\bz_\indexD$ and thus $\bz_{\indObs}$. Conditional on $\zobs$, the missing dimensions $\zmis$ follows normal distribution with mean $\Erm[\bz_\indMis| \zobs,\Sigma]=\Sigma_{\indMis,\indObs}\Sigma_{\indObs,\indObs}^{-1} \bz_{\indObs}$. Further taking expectation of $\zobs$ conditional on observation, we obtain
\begin{align}
\Erm[\bz_\indMis| \xobs,\Sigma] &=\Erm\left[\Erm[\bz_\indMis|\zobs, \Sigma]\big| \xobs,\Sigma\right]= \Sigma_{\indMis,\indObs}\Sigma_{\indObs,\indObs}^{-1} \Erm\left[\bz_{\indObs}| \xobs,\Sigma\right].
\label{Eq:meanMissing}\nonumber
\end{align}

One can compute $\Covrm[\bz_\indMis| \xobs,\Sigma]$ and  $\Covrm[\bz_\indMis,\bz_\indexD| \xobs,\Sigma]$ similarly: deferring details to the supplement, we find
\begin{equation*}
	\Covrm[\bz_\indMis,\bz_{\indObs}| \xobs,\Sigma]=  \Sigma_{\indMis,\indObs}\Sigma_{\indObs,\indObs}^{-1} \Covrm[\bz_{\indObs}| \xobs,\Sigma],
	\label{Eq:IncompleteMisObsCov}
	\end{equation*}
	\begin{multline}
	\Covrm[\bz_\indMis| \xobs,\Sigma]=\Sigma_{\indMis,\indMis}- \Sigma_{\indMis,\indObs}\Sigma_{\indObs,\indObs}^{-1} \Sigma_{\indObs,\indMis}\\ +\Sigma_{\indMis,\indObs}\Sigma_{\indObs,\indObs}^{-1} \Covrm[\bz_{\indObs}| \xobs,\Sigma] \Sigma_{\indObs,\indObs}^{-1} \Sigma_{\indObs,\indMis}, \nonumber
	\end{multline}
		where $\Covrm[\bz_{\indObs}| \xobs,\Sigma]$ has $\Covrm[\bz_{\indexD}| \xobs,\Sigma]$ as its submatrix and $0$ elsewhere, $\Covrm[\bz_\indMis,\bz_{\indObs}| \xobs,\Sigma]$ has $\Covrm[\bz_\indMis,\bz_\indexD| \xobs,\Sigma]$ as its submatrix and $0$ elsewhere.

\subsection{Approximating Truncated Normal Moments}
\label{sec:comp_ordinal}
Now it remains to compute $\Erm[\bz_{\indexD}| \xobs,\Sigma]$ and $\Covrm[\bz_{\indexD}| \xobs,\Sigma]$, which are the mean and covariance of a $|\indexD|$-dimensional normal truncated to $ \bigf_\indexD^{-1}(\bx_\indexD)$, a Cartesian product of intervals.
The computation involves multiple integrals of a nonlinear function and only admits a closed form expression when $|\indexD|=1$.
Direct computational methods \cite{bg2009moments}
are very expensive and can be inaccurate even for moderate $|\indexD|$. 
Notice the computation needs to be done for each row $\xio$ at each EM iteration separately,
thus sampling truncated normal distribution to evaluate the empirical moments \cite{pakman2014exact} is still expensive for large number of data points $n$.
Instead, we use a fast iterative method that scales well to large datasets,
following \cite{guo2015graphical}.

Suppose all but one element of $\bz_\indexD$ is known.
Then we can easily compute the resulting one dimensional truncated normal mean:
for $j\in \indexD$, if $\bz_j$ is unknown and $\bz_{\indexD-j}$ is known,
let $\Erm[z_j|\bz_{\indexD-j}, \xobs,\Sigma]=:g_j(\bz_{\indexD-j};x_j,\Sigma)$ define the nonlinear function $g_j: \mathbb{R}^{|\indexD|-1}\rightarrow  \mathbb{R}$,
parameterized by $x_j$ and $\Sigma$, detailed in the supplement.
We may also use $g_j$ to estimate $\Erm[z_j|\xobs,\Sigma]$
if $\Erm[\bz_{\indexD-j}| \xobs, \Sigma]$ is known:
\begin{align}
	&\Erm[z_j|\xobs,\Sigma] =\Erm[ \Erm[z_j|\bz_{\indexD-j}, \xobs,\Sigma]|\xobs, \Sigma ] \nonumber \\
	=&\Erm[ g_j(\bz_{\indexD-j};x_j,\Sigma)|\xobs, \Sigma ]\approx g_j(\Erm[\bz_{\indexD-j}| \xobs, \Sigma]; x_j,\Sigma), \label{Eq:approx_tmean}
\end{align}
if $g_j$ is approximately linear.
In other words, we can iteratively update the marginal mean of $\Erm[\bz_{\indexD}|\bx_{\indexO},\Sigma]$.
At EM iteration $t+1$, 
we conduct one iteration update with initial value from last EM iteration $\hat \bz_\indexD^{(t)} \approx \Erm[\bz_\indexD|\xobs,\Sigma^{(t)}]$:
\begin{equation}
\Erm[z_j|\xobs,\Sigma^{(t+1)}] \approx \hat z_j^{(t+1)} := g_j(\hat \bz_{\indexD-j}^{(t)}; x_j,\Sigma^{(t+1)}).
\label{Eq:meanRecursive}
\end{equation}
Surprisingly, one iteration update works well and more iterations do not bring significant improvement.

We use a diagonal approximation for $\Covrm \left[\bz_\indexD| \xobs,\Sigma\right]$:
we approximate $\Covrm \left[z_j, z_k| \xobs,\Sigma\right]$ as $0$
for $j \ne k\in \indexD$.
This approximation performs well when $z_j$ and $z_k$ are nearly independent given all observed information.
We approximate the diagonal entries\\ $\Varrm \left[z_j| \xobs,\Sigma^{(t+1)}\right]$ for $j\in\indexD$ using
a recursion similar to \cref{Eq:meanRecursive}, detailed in the supplement.

We point out the estimated covariance matrix in Eq. (\ref{Eq:G-func}) is the sum of the sample covariance matrix of the imputed $\zi$ using its conditional mean  
and the expected covariance brought by the imputation. 
The diagonal approximation only applies to the second term, while the first term is dense. Consequently, the estimator in Eq. (\ref{Eq:G-func}) is dense and can fit a large range of covariance matrices. 
Empirical evidence indicates that our approximation even outperforms the MCMC algorithm without such diagonal approximation \cite{hoff2007extending}, shown in Section \ref{sec:synthetic}.

\subsection{Computation Cost}
The complexity of each EM iteration is $O(\alpha np^3)$ with observed entry ratio $\alpha$.
The overall complexity is $O(T\alpha np^3)$, where $T$ is the number of EM steps required for convergence.
We found $T\leq 50$ in most of our experiments. On a laptop with Intel-i5-3.1GHz Core and 8 GB RAM, it takes $1.2$min for our algorithm to converge on a dataset with size $2000\times 60$ and $25\%$ missing entries (generated as in Section \ref{sec:synthetic} when $p=60$). 
Scaling our algorithm to large $p$ is important future work.
However, our algorithm is usually faster than many start-of-the-art imputation algorithms for large $n$ small $p$. Speed comparison on a dataset with size $6039\times 207$ is shown in Section \ref{sec:movie}.

	\section{EXPERIMENTS}
	Our first experiment demonstrates that our method, \texttt{Copula-EM}, is able to estimate a well-specified Gaussian copula model faster than the MCMC method \verb+sbgcop+ \cite{hoff2007extending, hoff2018package}.
	Our other experiments compare the accuracy of imputations produced by \texttt{Copula-EM} with
	\texttt{missForest} \cite{stekhoven2011missforest}, \texttt{xPCA} \cite{anderson2018xpca} and \texttt{imputeFAMD} \cite{audigier2016principal}, state-of-the-art nonparametric imputation algorithms for mixed data;
	and the low rank matrix completion algorithms \verb+softImpute+ \cite{mazumder2010spectral}
	and \texttt{GLRM} \cite{udell2016generalized}, which scale to large datasets. 
	\texttt{missForest} is implemented with recommended default settings: $10$ maximum iterations and $100$ trees  \cite{stekhoven2011using}.
	All other methods require selecting either the rank or the penalization parameter.
	We select them 	through 5-fold cross validation (5CV), unless otherwise specified.
	See the supplement for implementation details.
	For real datasets, we report results from our \texttt{Copula-EM} but put that from \texttt{sbgcop} in the supplement,
	since \texttt{Copula-EM} outperforms on all evaluation metrics
	and converges substantially faster. 

	To measure the imputation error on columns in $I$, we define a scaled mean absolute error (SMAE): 
	\[\textup{SMAE}:=\frac{1}{|I|}\sum_{j\in I}\frac{||\hat \bX_j - \bX_j||_1}{||\bX^{\textup{med}}_j - \bX_j||_1},
	\]
	where $\hat \bX_j, \bX^{\textup{med}}_j $ are the imputed values and observed median for $j$-th column, respectively. The estimator's SMAE is smaller than $1$ if it outperforms column median imputation. For each data type, the SMAE can be computed on corresponding columns. 
	To evaluate the estimated correlation, we use relative error $||\hat\Sigma-\Sigma||_{F}/||\Sigma||_{F}$, where $\hat\Sigma$ is the estimated correlation matrix.


	\subsection{Synthetic Data}
	\label{sec:synthetic}
	The first experiment compares the speed of the two algorithms to estimate Gaussian copula models: \texttt{Copula-EM} and \texttt{sbgcop}. 
	Note \texttt{Copula-EM} is implemented in pure R, 
	while the computational core of \texttt{sbgcop} is implemented in C. Hence further acceleration of \texttt{Copula-EM} is possible. 


	We generate 100 synthetic datasets with $n=2000$ observations and $p=15$ variables
	from a well-specified Gaussian copula model with random $\Sigma$ generated \cite{qiu2009clustergeneration}.
	For each $\Sigma$,
	first generate rows of $\bZ\in \mathbb{R}^{n\times p}$ as $\bz^1,\cdots,\bz^{n}\overset{i.i.d.}{\sim} \mathcal{N}(\bo,\Sigma)$. Then generate $\bX=\bigf(\bZ)$ using monotone $\bigf$ such that $\bX_1,\ldots,\bX_5$ have exponential distributions, $\bX_6,\ldots,\bX_{10}$ are binary and  $\bX_{11},\ldots,\bX_{15}$ are $1$-$5$ ordinal. 
	
\begin{figure}[h]
		\centering
		\includegraphics[width=\linewidth]{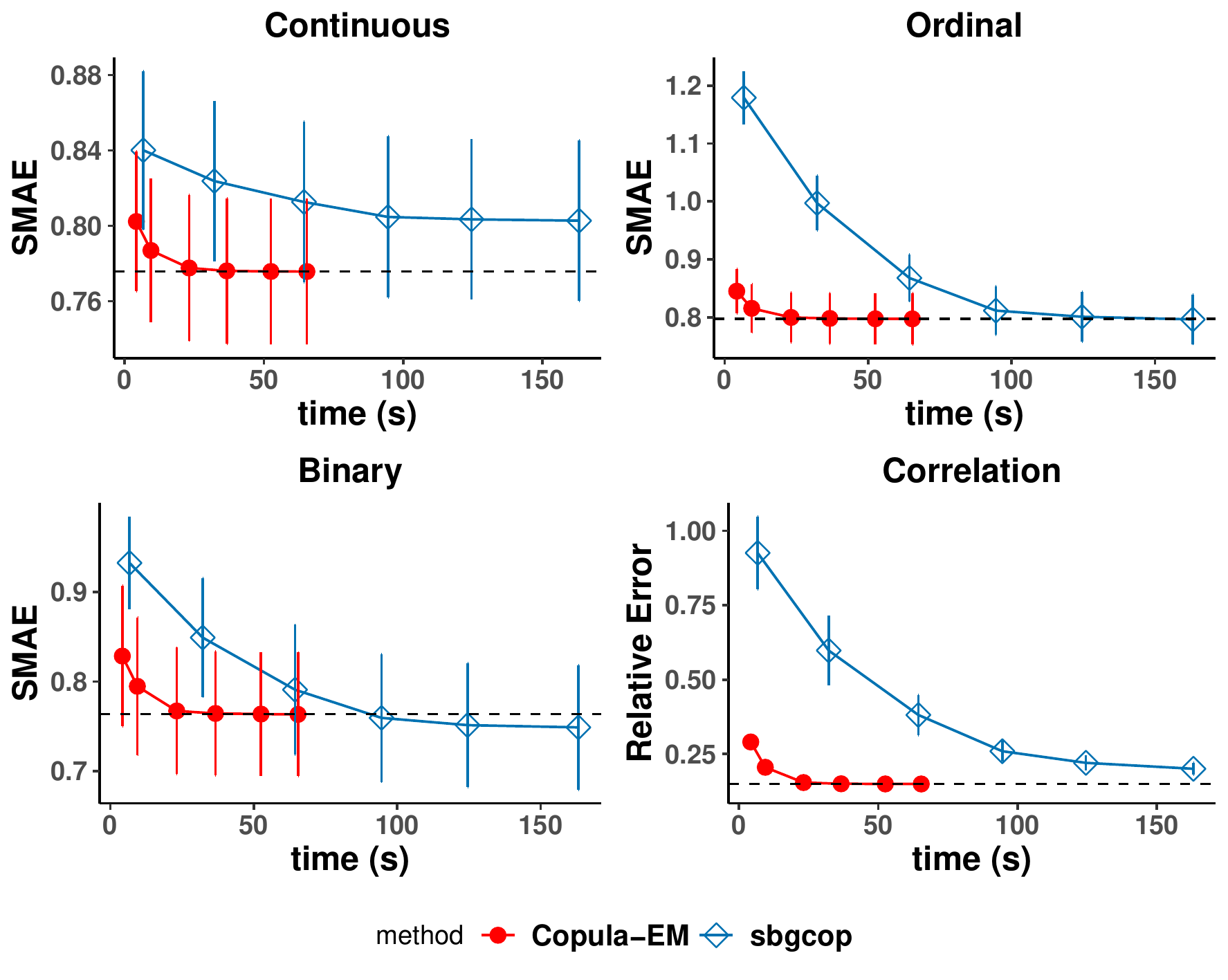}
		\caption{\texttt{Copula-EM} vs \texttt{sbgcop}: The imputation error for each data type and estimated correlation error over time cost. Dashed line indicates the final error of \texttt{Copula-EM}.}
		\label{fig:sim2}
	\end{figure}

	We randomly remove $30\%$ of the entries of $\bX$,
	train \texttt{Copula-EM} and \texttt{sbgcop}, 
	and compute the imputation error on the held-out set.
	We plot the imputation accuracy and correlation estimation accuracy versus runtime of each algorithm in Figure \ref{fig:sim2}. 
	\texttt{Copula-EM} converges quickly, in about 25s,
	while \texttt{sbgcop} takes much longer and suffers 
	high error at shorter times.
    \texttt{Copula-EM} estimates correlations and continuous imputations at convergence more accurately than \texttt{sbgcop} even when the latter algorithm is given $6$ times more runtime. 
	Interestingly, \texttt{Copula-EM} recovers the correlation matrix better than \texttt{sbgcop} even asymptotically. 
	These results demonstrate the impact of the approximate EM algorithm \ref{sec:comp_ordinal} compared to the (fully accurate) MCMC model of \texttt{sbgcop}: the approximation allows faster convergence, to an estimate of nearly the same quality.

    For ordinal data imputation, \texttt{Copula-EM} reaches the same performance as \texttt{sbgcop} $6$ times faster.
	For binary data imputation, \texttt{sbgcop} is four times slower than \texttt{Copula-EM} at reaching the final performance of \texttt{Copula-EM}, but \texttt{sbgcop} outperforms \texttt{Copula-EM} given even more time.  
We conjecture that the drop in imputation accuracy of \texttt{Copula-EM} for binary data could be mitigated using multiple imputation \cite[Chapter~5.4]{little2019statistical}, as outlined in Sec \ref{sec:imp} by combining the imputations (using mean or median) into a single imputation to reduce the effect of approximating the truncated normal distribution. 

\begin{figure}[h]
		\centering
		\includegraphics[width=\linewidth]{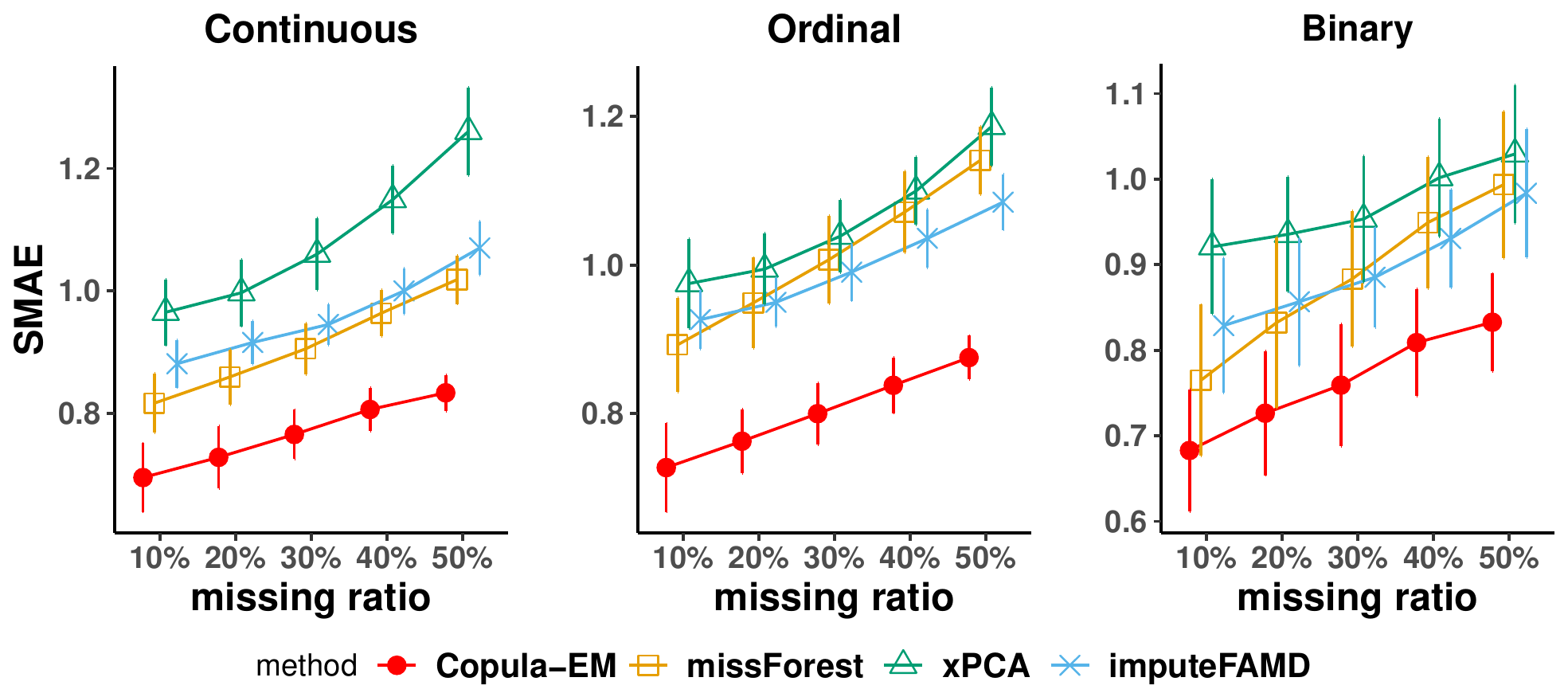}
		\caption{\texttt{Copula-EM} vs nonparametric algorithms: The imputation error for each data type on synthetic data.}
		\label{fig:more_sim2}
	\end{figure}

	The second experiment compares the imputation accuracy of \texttt{Copula-EM} and nonparametric algorithms.
	Using the same data generation mechanism, we randomly remove $10\% - 50\%$ of  the entries of $\bX$.
	The optimal rank selected using 5CV is $3$ for \texttt{xPCA} and $6$ for \texttt{imputeFAMD}.
	Shown in Figure \ref{fig:more_sim2},\texttt{Copula-EM} substantially outperforms all nonparametric algorithms for all data types.

	\subsection{General Social Survey (GSS) Data}
\begin{figure}[h]
		\centering
		\includegraphics[width=0.7\linewidth]{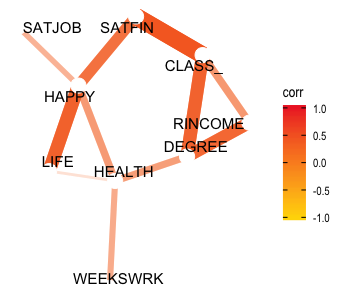}
		\caption{High Correlations $(|\cdot|>0.3)$ of $5$ interesting variables from GSS data are plotted.}
		\label{fig:GSS}
	\end{figure}

	We chose $18$ variables with $2538$ observations from GSS dataset in year 2014. $24.9\%$ of the entries are missing. The dataset consists of $1$ continuous (\texttt{AGE}) and $17$ ordinal variables with 2 to 48 levels. We investigate the imputation accuracy on five selected variables: \texttt{INCOME}, \texttt{LIFE}, \texttt{HEALTH}, \texttt{CLASS}\footnote{Subjective class identification from lower to upper class} and \texttt{HAPPY}. For each variable, we sample $1500$ observation and divide them into $20$ folds. We mask one fold of only one variable as test data in each experiment. 
	The selected rank is $2$ for both \texttt{xPCA} and \texttt{imputeFAMD}.
	We report the SMAE for each variable in Table \ref{table:gss}. Our method performs the best for all variables. Further our method always performs better than median imputation. In contrast, the other three methods perform worse than median imputation for some variables. Our method also provides estimated variable correlation, which is usually desired in social survey study. We plot high correlations from the copula correlation matrix as a graph in Figure \ref{fig:GSS}. 
	\begin{table}
		\centering
		\caption{Imputation Error on Five GSS Variables}
		\begin{tabular}{ccccc}
			\toprule
			Variable & Copula-EM & missForest & xPCA &imputeFAMD\\ \midrule
			CLASS	& \boldmath$0.735(0.10)$ &  $0.782(0.09)$ & $0.795(0.08)$& $0.797(0.10)$ \\
			LIFE & \boldmath$0.759(0.12)$ &  $0.828(0.17)$& $0.783(0.11)$ & $0.821(0.11)$ \\
			HEALTH & \boldmath$0.877(0.09)$ & $1.143(0.18)$ & $0.908(0.10)$ & $0.947(0.04)$\\
			HAPPY &  \boldmath$0.896(0.08)$ &  $1.079(0.15)$& $1.003(0.15)$ & $1.001(0.10)$\\
			INCOME & \boldmath$0.869(0.07)$ & $0.944(0.18)$& $1.090(0.15)$ & $0.996(0.01)$\\
			\bottomrule
		\end{tabular}
		\label{table:gss}
	\end{table}
	
\subsection{MovieLens 1M Data}
\label{sec:movie}
Recall our method scales cubicly in the number of variables. 
Hence for this experiment, we sample the subset of the MovieLens 1M data \cite{harper2016movielens} consisting of the $207$ movies with at least $1000$ ratings and all users who rate at least one of those $207$ movies. On this subset, $75.6\%$ of entries are missing. 
Under the time limit $1$ hour,
we implement all algorithms but \texttt{imputeFAMD}.
\texttt{Copula-EM} takes 9 mins and \texttt{missForest} takes 25 mins. 
These two methods have no parameters to tune.
To select tuning parameters for other algorithms,
we manually mask $10\%$ of the data for the test set and use the remaining data to train the model, and repeat 20 times.
The selected rank using 5CV is $99$ for \texttt{softImpute}, $6$ for \texttt{xPCA} and $8$ for \texttt{GLRM} with bigger-vs-smaller loss. 
With the selected tuning parameter, 
low rank matrix completion methods are substantially faster. 
For example, \texttt{softImpute} only takes 33s. 
However, counting the additional time to select tuning parameters using 5CV,
\texttt{softImpute} takes 16mins to select the penalization parameter with regularization path length 50,
which is already more expensive than \texttt{Copula-EM}. 
Interestingly, the ranks selected are quite different even when the models perform similarly: \texttt{GLRM} chooses rank $8$ while \texttt{softImpute} chooses rank $99$.

We report both mean absolute error (MAE) and RMSE in Table \ref{table:movie}. Our method outperforms all others in both MAE and RMSE.
This result is notable, because \texttt{Copula-EM} does not directly minimize MAE or RMSE, while \texttt{softImpute} directly minimizes RMSE.
It also indicates \texttt{Copula-EM} does not overfit even with $O(p^2)$ free parameters.

	\begin{table}
		\caption{Imputation Error on 207 Movies}
		\centering
		\begin{tabular}{lll}
			\toprule
			Algorithm     & MAE     & RMSE \\
			\midrule
			Column Median  & $0.702(0.004)$       & $1.001(0.004)$    \\
			Copula-EM & \boldmath $0.579(0.004)$   & \boldmath$0.880(0.005)$ \\
			GLRM   &  $0.595(0.004)$   &    $0.892(0.004)$  \\
			softImpute   &  $0.602(0.004)$   &    $0.883(0.004)$  \\
			xPCA & $0.613(0.004)$ & $0.897(0.004)$ \\
			missForest  & $0.669(0.004)$        & $1.015(0.006)$   \\
			\bottomrule
		\end{tabular}
		\label{table:movie}
	\end{table}
	
	\subsection{Music Auto-tagging: CAL500exp Data}
	The CAL500 expansion (CAL500exp) dataset 
	\cite{wang2014towards} is an enriched version of the well-known CAL500 dataset \cite{turnbull2007towards}. This dataset consists of 67 binary tags (including genre, mood and instrument, labeled by experts) to $3223$ music fragments from $500$ songs. 
	Music auto-tagging is a multi-label learning problem. 
	A feature vector is usually computed first based on the music files and then a classifier is trained for each tag. This procedure is expensive and neglects the association among known labels. We treat this task as a missing data imputation problem and only use observed labels to impute unknown labels. This dataset is completely observed. 
	We randomly remove some portions of the observed labels as a test set and repeat $20$ times.
	The selected optimal rank is $4$ for \texttt{xPCA} and $15$ for \texttt{imputeFAMD}.
	Shown in \cref{table:cal500}, \texttt{Copula-EM} performs the best in terms of SMAE. 
	The superiority of \texttt{Copula-EM} over other algorithms substantially grows as the missing ratio increases. 
	Moreover, \texttt{Copula-EM} yields very stable imputations: the standard deviation of its SMAE is imperceptibly small.
	
	
		\begin{table}
		\caption{Imputation Error (SMAE) on CAL500exp.}
		\centering
		\begin{tabular}{llll}
			\toprule
			Algorithm     & $40\%$ missing     & $50\%$ missing  & $60\%$ missing   \\
			\midrule
			Copula-EM & \boldmath $0.799(0.002)$  & \boldmath$0.822(0.003)$ & \boldmath$0.849(0.002)$ \\
			missForest  & $0.800(0.018)$ & $0.984(0.026)$& $1.181(0.024)$  \\
			imputeFAMD & $0.823(0.013)$ &  $0.920(0.016)$ & $1.114(0.020)$\\
			xPCA & $0.911(0.018)$ & $0.988(0.071)$ & $1.108(0.145)$ \\
			\bottomrule
		\end{tabular}
		\label{table:cal500}
	\end{table}

	\subsection{More Ordinal Data and Mixed Data}
	\label{sec:moredata}
	We compare mixed data imputation algorithms on two more ordinal classification datasets\footnote{Available at https://waikato.github.io/weka-wiki/datasets/}, Lecturers Evaluation \textit{(LEV)} and Employee Selection \textit{(ESL)}, and two more mixed datasets, German Breast Cancer Study Group \textit{(GBSG)}\footnote{Available at https://cran.r-project.org/web/packages/mfp/} and Restaurant Tips \textit{(TIPS)}\footnote{Available at http://ggobi.org/book/}. Dataset descriptions appear in Table \ref{table:extra}, and more details appear in the supplement. All datasets are completely observed.

For each dataset, we randomly remove $30\%$ entries as a test set and repeat $100$ times. For ordinal classification datasets, we evaluate the SMAE for the label and for the features, respectively. For mixed datasets, we evaluate the SMAE for ordinal dimensions and for continuous dimensions, respectively. We report results in Table \ref{table:extra}. Our method outperforms the others in all but one setting, often by a substantial margin.

		\begin{table*}
		\centering
		\caption{Imputation Error on More Ordinal and Mixed Datasets.}
		\begin{tabular}{l|c|c|lccccc}
			\toprule
	 Dataset & Size & Selected Rank & Type	 & Copula-EM & missForest & xPCA & imputeFAMD\\ \midrule
	ESL& $488\times 5$ 
		& 1 (xPCA)
		& Label & \boldmath$0.372(0.04)$ & $0.553(0.08)$ & $0.404(0.04)$ & $0.503(0.06)$\\
	& 4 features, 1 label 
	    & 5 (imputeFAMD)
	    & Feature  & \boldmath$0.584(0.03)$ &  $0.873(0.06)$& $0.668(0.03)$& $0.687(0.03)$\\
		\hline
	LEV& $1000\times 5$ 
	    & 1 (xPCA)
	    &Label  &\boldmath$0.750(0.04)$ & $0.970(0.09)$ & $0.860(0.06)$ & $0.882(0.05)$\\
	&  4 features, 1 label &
	    5 (imputeFAMD) &
		Feature & $0.907(0.01)$  & \boldmath  $0.799(0.03)$ & $1.037(0.02)$ & $1.085(0.04)$\\
		\hline
	GBSG& $686\times 10$ & 
	    2 (xPCA) &
		Ordinal  &\boldmath$0.793(0.03)$ &  $0.887(0.05)$ & $0.876(0.04)$ & $0.840(0.03)$\\
	&6 continuous, 4 ordinal &
	    2 (imputeFAMD)&
	    Continuous  &\boldmath$0.876(0.01)$ &  $1.029(0.03)$ & $1.100(0.04)$ & $1.038(0.03)$\\
		\hline
	TIPS& $244\times 7$ &
	    2 (xPCA) &
	    Ordinal &\boldmath$0.786(0.05)$ &  $0.928(0.09)$ & $0.928(0.08)$ & $0.891(0.09)$\\
	& 2 continuous, 5 ordinal  &
	    6 (imputeFAMD) &
	    Continuous & \boldmath$0.755(0.04)$ & $0.837(0.05)$ & $1.011(0.11)$ & $0.892(0.13)$\\
			\bottomrule
		\end{tabular}
		\label{table:extra}
	\end{table*}

	\section{SUMMARY AND DISCUSSION}
	In this paper, we proposed an imputation algorithm that models mixed data with a Gaussian copula model,
	together with an effective approximate EM algorithm to estimate the copula correlation with incomplete mixed data.
	Our algorithm has no tuning parameter and are easy to implement.
	Our experiments demonstrate the success of the proposed method.
	Scaling these methods to larger datasets (especially, with more columns),
	constitutes important future work.

	We end by noting a few contrasts between the present approach
	and typical low rank approximation methods for data imputation.
	Low rank approximation constructs a latent simple (low rank) object
	and posits that observations are noisy draws from that simple latent object.
	In contrast, our approach uses a parametric, but full-dimensional, model for
	the latent object; observations are given by a deterministic function of the latent object.
	In other words, in previous work the latent object is exact and the observations are noisy;
	in our work, the latent object is noisy and the observations are exact.
	Which more faithfully models real data?
	As evidence, we might consider whether low rank models agree on the best
	rank to fit a given dataset.
	For example, on the MovieLens dataset:
	(1) The low rank matrix completion methods \texttt{xPCA} and \texttt{GLRM},
	implemented using alternating minimization, select small optimal ranks (6 and 8),
	while \texttt{softImpute}, implemented using nuclear norm minimization,
	selects the much larger optimal rank $99$.
	(2) Our algorithm outperforms all the low rank matrix completion methods we tested.
	These observations suggest the  low rank assumption commonly used to
	fit the MovieLens dataset may not be fundamental,
	but may arise as a mathematical artifact \cite{udell2019why}.
	More supporting empirical results can be found in \cite{avron2012efficient}: the performance of \texttt{softImpute} keeps improving as the rank increases (up to $10^3$).

\begin{acks}
We gratefully acknowledge support from
NSF Awards IIS-1943131 
and CCF-1740822, 
the ONR Young Investigator Program, %
DARPA Award FA8750-17-2-0101, 
the Simons Institute, 
Canadian Institutes of Health Research, 
and Capital One.
We thank Clifford Anderson-Bergman and Tamara G. Kolda for help in understanding and implementing xPCA, Julie Josse for help in implementing imputeFAMD, and Yang Ning and Zhengze Zhou for helpful discussions.
Special thanks go to Xiaoyi Zhu for help in producing Figure 3.
\end{acks}

\bibliographystyle{ACM-Reference-Format}
\bibliography{impute}


\begin{thebibliography}{36}


\ifx \showCODEN    \undefined \def \showCODEN     #1{\unskip}     \fi
\ifx \showDOI      \undefined \def \showDOI       #1{#1}\fi
\ifx \showISBNx    \undefined \def \showISBNx     #1{\unskip}     \fi
\ifx \showISBNxiii \undefined \def \showISBNxiii  #1{\unskip}     \fi
\ifx \showISSN     \undefined \def \showISSN      #1{\unskip}     \fi
\ifx \showLCCN     \undefined \def \showLCCN      #1{\unskip}     \fi
\ifx \shownote     \undefined \def \shownote      #1{#1}          \fi
\ifx \showarticletitle \undefined \def \showarticletitle #1{#1}   \fi
\ifx \showURL      \undefined \def \showURL       {\relax}        \fi
\providecommand\bibfield[2]{#2}
\providecommand\bibinfo[2]{#2}
\providecommand\natexlab[1]{#1}
\providecommand\showeprint[2][]{arXiv:#2}

\bibitem[\protect\citeauthoryear{Anderson-Bergman, Kolda, and
  Kincher-Winoto}{Anderson-Bergman et~al\mbox{.}}{2018}]%
        {anderson2018xpca}
\bibfield{author}{\bibinfo{person}{Clifford Anderson-Bergman},
  \bibinfo{person}{Tamara~G Kolda}, {and} \bibinfo{person}{Kina
  Kincher-Winoto}.} \bibinfo{year}{2018}\natexlab{}.
\newblock \showarticletitle{{XPCA}: Extending {PCA} for a Combination of
  Discrete and Continuous Variables}.
\newblock \bibinfo{journal}{\emph{arXiv preprint arXiv:1808.07510}}
  (\bibinfo{year}{2018}).
\newblock


\bibitem[\protect\citeauthoryear{Audigier, Husson, and Josse}{Audigier
  et~al\mbox{.}}{2016}]%
        {audigier2016principal}
\bibfield{author}{\bibinfo{person}{Vincent Audigier},
  \bibinfo{person}{Fran{\c{c}}ois Husson}, {and} \bibinfo{person}{Julie
  Josse}.} \bibinfo{year}{2016}\natexlab{}.
\newblock \showarticletitle{A principal component method to impute missing
  values for mixed data}.
\newblock \bibinfo{journal}{\emph{Advances in Data Analysis and
  Classification}} \bibinfo{volume}{10}, \bibinfo{number}{1}
  (\bibinfo{year}{2016}), \bibinfo{pages}{5--26}.
\newblock


\bibitem[\protect\citeauthoryear{Avron, Kale, Kasiviswanathan, and
  Sindhwani}{Avron et~al\mbox{.}}{2012}]%
        {avron2012efficient}
\bibfield{author}{\bibinfo{person}{Haim Avron}, \bibinfo{person}{Satyen Kale},
  \bibinfo{person}{Shiva Kasiviswanathan}, {and} \bibinfo{person}{Vikas
  Sindhwani}.} \bibinfo{year}{2012}\natexlab{}.
\newblock \showarticletitle{Efficient and practical stochastic subgradient
  descent for nuclear norm regularization}.
\newblock \bibinfo{journal}{\emph{arXiv preprint arXiv:1206.6384}}
  (\bibinfo{year}{2012}).
\newblock


\bibitem[\protect\citeauthoryear{BG and Wilhelm}{BG and Wilhelm}{2009}]%
        {bg2009moments}
\bibfield{author}{\bibinfo{person}{Manjunath BG} {and} \bibinfo{person}{Stefan
  Wilhelm}.} \bibinfo{year}{2009}\natexlab{}.
\newblock \showarticletitle{Moments calculation for the double truncated
  multivariate normal density}.
\newblock \bibinfo{journal}{\emph{Available at SSRN 1472153}}
  (\bibinfo{year}{2009}).
\newblock


\bibitem[\protect\citeauthoryear{Cand{\`e}s and Recht}{Cand{\`e}s and
  Recht}{2009}]%
        {candes2009exact}
\bibfield{author}{\bibinfo{person}{Emmanuel~J Cand{\`e}s} {and}
  \bibinfo{person}{Benjamin Recht}.} \bibinfo{year}{2009}\natexlab{}.
\newblock \showarticletitle{Exact matrix completion via convex optimization}.
\newblock \bibinfo{journal}{\emph{Foundations of Computational mathematics}}
  \bibinfo{volume}{9}, \bibinfo{number}{6} (\bibinfo{year}{2009}),
  \bibinfo{pages}{717}.
\newblock


\bibitem[\protect\citeauthoryear{Cui, Bucur, Groot, and Heskes}{Cui
  et~al\mbox{.}}{2019}]%
        {cui2019novel}
\bibfield{author}{\bibinfo{person}{Ruifei Cui}, \bibinfo{person}{Ioan~Gabriel
  Bucur}, \bibinfo{person}{Perry Groot}, {and} \bibinfo{person}{Tom Heskes}.}
  \bibinfo{year}{2019}\natexlab{}.
\newblock \showarticletitle{A novel Bayesian approach for latent variable
  modeling from mixed data with missing values}.
\newblock \bibinfo{journal}{\emph{Statistics and Computing}}
  \bibinfo{volume}{29}, \bibinfo{number}{5} (\bibinfo{year}{2019}),
  \bibinfo{pages}{977--993}.
\newblock


\bibitem[\protect\citeauthoryear{Dvoretzky, Kiefer, Wolfowitz,
  et~al\mbox{.}}{Dvoretzky et~al\mbox{.}}{1956}]%
        {dvoretzky1956asymptotic}
\bibfield{author}{\bibinfo{person}{Aryeh Dvoretzky}, \bibinfo{person}{Jack
  Kiefer}, \bibinfo{person}{Jacob Wolfowitz}, {et~al\mbox{.}}}
  \bibinfo{year}{1956}\natexlab{}.
\newblock \showarticletitle{Asymptotic minimax character of the sample
  distribution function and of the classical multinomial estimator}.
\newblock \bibinfo{journal}{\emph{The Annals of Mathematical Statistics}}
  \bibinfo{volume}{27}, \bibinfo{number}{3} (\bibinfo{year}{1956}),
  \bibinfo{pages}{642--669}.
\newblock


\bibitem[\protect\citeauthoryear{Fan, Liu, Ning, and Zou}{Fan
  et~al\mbox{.}}{2017}]%
        {fan2017high}
\bibfield{author}{\bibinfo{person}{Jianqing Fan}, \bibinfo{person}{Han Liu},
  \bibinfo{person}{Yang Ning}, {and} \bibinfo{person}{Hui Zou}.}
  \bibinfo{year}{2017}\natexlab{}.
\newblock \showarticletitle{High dimensional semiparametric latent graphical
  model for mixed data}.
\newblock \bibinfo{journal}{\emph{Journal of the Royal Statistical Society:
  Series B (Statistical Methodology)}} \bibinfo{volume}{79},
  \bibinfo{number}{2} (\bibinfo{year}{2017}), \bibinfo{pages}{405--421}.
\newblock


\bibitem[\protect\citeauthoryear{Feng and Ning}{Feng and Ning}{2019}]%
        {feng2019high}
\bibfield{author}{\bibinfo{person}{Huijie Feng} {and} \bibinfo{person}{Yang
  Ning}.} \bibinfo{year}{2019}\natexlab{}.
\newblock \showarticletitle{High-dimensional Mixed Graphical Model with Ordinal
  Data: Parameter Estimation and Statistical Inference}. In
  \bibinfo{booktitle}{\emph{The 22nd International Conference on Artificial
  Intelligence and Statistics}}. \bibinfo{pages}{654--663}.
\newblock


\bibitem[\protect\citeauthoryear{Ganti, Balzano, and Willett}{Ganti
  et~al\mbox{.}}{2015}]%
        {ganti2015matrix}
\bibfield{author}{\bibinfo{person}{Ravi~Sastry Ganti}, \bibinfo{person}{Laura
  Balzano}, {and} \bibinfo{person}{Rebecca Willett}.}
  \bibinfo{year}{2015}\natexlab{}.
\newblock \showarticletitle{Matrix completion under monotonic single index
  models}. In \bibinfo{booktitle}{\emph{Advances in Neural Information
  Processing Systems}}. \bibinfo{pages}{1873--1881}.
\newblock


\bibitem[\protect\citeauthoryear{Guo, Levina, Michailidis, and Zhu}{Guo
  et~al\mbox{.}}{2015}]%
        {guo2015graphical}
\bibfield{author}{\bibinfo{person}{Jian Guo}, \bibinfo{person}{Elizaveta
  Levina}, \bibinfo{person}{George Michailidis}, {and} \bibinfo{person}{Ji
  Zhu}.} \bibinfo{year}{2015}\natexlab{}.
\newblock \showarticletitle{Graphical models for ordinal data}.
\newblock \bibinfo{journal}{\emph{Journal of Computational and Graphical
  Statistics}} \bibinfo{volume}{24}, \bibinfo{number}{1}
  (\bibinfo{year}{2015}), \bibinfo{pages}{183--204}.
\newblock


\bibitem[\protect\citeauthoryear{Harper and Konstan}{Harper and
  Konstan}{2016}]%
        {harper2016movielens}
\bibfield{author}{\bibinfo{person}{F~Maxwell Harper} {and}
  \bibinfo{person}{Joseph~A Konstan}.} \bibinfo{year}{2016}\natexlab{}.
\newblock \showarticletitle{The movielens datasets: History and context}.
\newblock \bibinfo{journal}{\emph{Acm transactions on interactive intelligent
  systems (tiis)}} \bibinfo{volume}{5}, \bibinfo{number}{4}
  (\bibinfo{year}{2016}), \bibinfo{pages}{19}.
\newblock


\bibitem[\protect\citeauthoryear{Hoff and Hoff}{Hoff and Hoff}{2018}]%
        {hoff2018package}
\bibfield{author}{\bibinfo{person}{Peter Hoff} {and}
  \bibinfo{person}{Maintainer~Peter Hoff}.} \bibinfo{year}{2018}\natexlab{}.
\newblock \showarticletitle{Package ‘sbgcop’}.
\newblock  (\bibinfo{year}{2018}).
\newblock


\bibitem[\protect\citeauthoryear{Hoff et~al\mbox{.}}{Hoff
  et~al\mbox{.}}{2007}]%
        {hoff2007extending}
\bibfield{author}{\bibinfo{person}{Peter~D Hoff} {et~al\mbox{.}}}
  \bibinfo{year}{2007}\natexlab{}.
\newblock \showarticletitle{Extending the rank likelihood for semiparametric
  copula estimation}.
\newblock \bibinfo{journal}{\emph{The Annals of Applied Statistics}}
  \bibinfo{volume}{1}, \bibinfo{number}{1} (\bibinfo{year}{2007}),
  \bibinfo{pages}{265--283}.
\newblock


\bibitem[\protect\citeauthoryear{Hollenbach, Bojinov, Minhas, Metternich, Ward,
  and Volfovsky}{Hollenbach et~al\mbox{.}}{2018}]%
        {hollenbach2018multiple}
\bibfield{author}{\bibinfo{person}{Florian~M Hollenbach},
  \bibinfo{person}{Iavor Bojinov}, \bibinfo{person}{Shahryar Minhas},
  \bibinfo{person}{Nils~W Metternich}, \bibinfo{person}{Michael~D Ward}, {and}
  \bibinfo{person}{Alexander Volfovsky}.} \bibinfo{year}{2018}\natexlab{}.
\newblock \showarticletitle{Multiple Imputation Using Gaussian Copulas}.
\newblock \bibinfo{journal}{\emph{Sociological Methods \& Research}}
  (\bibinfo{year}{2018}), \bibinfo{pages}{0049124118799381}.
\newblock


\bibitem[\protect\citeauthoryear{Keshavan, Montanari, and Oh}{Keshavan
  et~al\mbox{.}}{2010}]%
        {keshavan2010matrix}
\bibfield{author}{\bibinfo{person}{Raghunandan~H Keshavan},
  \bibinfo{person}{Andrea Montanari}, {and} \bibinfo{person}{Sewoong Oh}.}
  \bibinfo{year}{2010}\natexlab{}.
\newblock \showarticletitle{Matrix completion from noisy entries}.
\newblock \bibinfo{journal}{\emph{Journal of Machine Learning Research}}
  \bibinfo{volume}{11}, \bibinfo{number}{Jul} (\bibinfo{year}{2010}),
  \bibinfo{pages}{2057--2078}.
\newblock


\bibitem[\protect\citeauthoryear{Kosorok}{Kosorok}{2008}]%
        {kosorok2008introduction}
\bibfield{author}{\bibinfo{person}{Michael~R Kosorok}.}
  \bibinfo{year}{2008}\natexlab{}.
\newblock \bibinfo{booktitle}{\emph{Introduction to empirical processes and
  semiparametric inference.}}
\newblock \bibinfo{publisher}{Springer}.
\newblock


\bibitem[\protect\citeauthoryear{Little and Rubin}{Little and Rubin}{2019}]%
        {little2019statistical}
\bibfield{author}{\bibinfo{person}{Roderick~JA Little} {and}
  \bibinfo{person}{Donald~B Rubin}.} \bibinfo{year}{2019}\natexlab{}.
\newblock \bibinfo{booktitle}{\emph{Statistical analysis with missing data}}.
  Vol.~\bibinfo{volume}{793}.
\newblock \bibinfo{publisher}{Wiley}.
\newblock


\bibitem[\protect\citeauthoryear{Liu, Lafferty, and Wasserman}{Liu
  et~al\mbox{.}}{2009}]%
        {liu2009nonparanormal}
\bibfield{author}{\bibinfo{person}{Han Liu}, \bibinfo{person}{John Lafferty},
  {and} \bibinfo{person}{Larry Wasserman}.} \bibinfo{year}{2009}\natexlab{}.
\newblock \showarticletitle{The nonparanormal: Semiparametric estimation of
  high dimensional undirected graphs}.
\newblock \bibinfo{journal}{\emph{Journal of Machine Learning Research}}
  \bibinfo{volume}{10}, \bibinfo{number}{Oct} (\bibinfo{year}{2009}),
  \bibinfo{pages}{2295--2328}.
\newblock


\bibitem[\protect\citeauthoryear{Mazumder, Hastie, and Tibshirani}{Mazumder
  et~al\mbox{.}}{2010}]%
        {mazumder2010spectral}
\bibfield{author}{\bibinfo{person}{Rahul Mazumder}, \bibinfo{person}{Trevor
  Hastie}, {and} \bibinfo{person}{Robert Tibshirani}.}
  \bibinfo{year}{2010}\natexlab{}.
\newblock \showarticletitle{Spectral regularization algorithms for learning
  large incomplete matrices}.
\newblock \bibinfo{journal}{\emph{Journal of machine learning research}}
  \bibinfo{volume}{11}, \bibinfo{number}{Aug} (\bibinfo{year}{2010}),
  \bibinfo{pages}{2287--2322}.
\newblock


\bibitem[\protect\citeauthoryear{McLachlan and Krishnan}{McLachlan and
  Krishnan}{2007}]%
        {mclachlan2007algorithm}
\bibfield{author}{\bibinfo{person}{Geoffrey McLachlan} {and}
  \bibinfo{person}{Thriyambakam Krishnan}.} \bibinfo{year}{2007}\natexlab{}.
\newblock \bibinfo{booktitle}{\emph{The EM algorithm and extensions}}.
  Vol.~\bibinfo{volume}{382}.
\newblock \bibinfo{publisher}{John Wiley \& Sons}.
\newblock


\bibitem[\protect\citeauthoryear{Murray, Dunson, Carin, and Lucas}{Murray
  et~al\mbox{.}}{2013}]%
        {murray2013bayesian}
\bibfield{author}{\bibinfo{person}{Jared~S Murray}, \bibinfo{person}{David~B
  Dunson}, \bibinfo{person}{Lawrence Carin}, {and} \bibinfo{person}{Joseph~E
  Lucas}.} \bibinfo{year}{2013}\natexlab{}.
\newblock \showarticletitle{Bayesian Gaussian copula factor models for mixed
  data}.
\newblock \bibinfo{journal}{\emph{J. Amer. Statist. Assoc.}}
  \bibinfo{volume}{108}, \bibinfo{number}{502} (\bibinfo{year}{2013}),
  \bibinfo{pages}{656--665}.
\newblock


\bibitem[\protect\citeauthoryear{Pakman and Paninski}{Pakman and
  Paninski}{2014}]%
        {pakman2014exact}
\bibfield{author}{\bibinfo{person}{Ari Pakman} {and} \bibinfo{person}{Liam
  Paninski}.} \bibinfo{year}{2014}\natexlab{}.
\newblock \showarticletitle{Exact hamiltonian monte carlo for truncated
  multivariate gaussians}.
\newblock \bibinfo{journal}{\emph{Journal of Computational and Graphical
  Statistics}} \bibinfo{volume}{23}, \bibinfo{number}{2}
  (\bibinfo{year}{2014}), \bibinfo{pages}{518--542}.
\newblock


\bibitem[\protect\citeauthoryear{Qiu and Joe}{Qiu and Joe}{2009}]%
        {qiu2009clustergeneration}
\bibfield{author}{\bibinfo{person}{Weiliang Qiu} {and} \bibinfo{person}{Harry
  Joe}.} \bibinfo{year}{2009}\natexlab{}.
\newblock \showarticletitle{clusterGeneration: random cluster generation (with
  specified degree of separation)}.
\newblock \bibinfo{journal}{\emph{R package version}} \bibinfo{volume}{1},
  \bibinfo{number}{7} (\bibinfo{year}{2009}), \bibinfo{pages}{75275--0122}.
\newblock


\bibitem[\protect\citeauthoryear{Recht, Fazel, and Parrilo}{Recht
  et~al\mbox{.}}{2010}]%
        {recht2010guaranteed}
\bibfield{author}{\bibinfo{person}{Benjamin Recht}, \bibinfo{person}{Maryam
  Fazel}, {and} \bibinfo{person}{Pablo~A Parrilo}.}
  \bibinfo{year}{2010}\natexlab{}.
\newblock \showarticletitle{Guaranteed minimum-rank solutions of linear matrix
  equations via nuclear norm minimization}.
\newblock \bibinfo{journal}{\emph{SIAM review}} \bibinfo{volume}{52},
  \bibinfo{number}{3} (\bibinfo{year}{2010}), \bibinfo{pages}{471--501}.
\newblock


\bibitem[\protect\citeauthoryear{Rennie and Srebro}{Rennie and Srebro}{2005a}]%
        {rennie2005fast}
\bibfield{author}{\bibinfo{person}{Jasson~DM Rennie} {and}
  \bibinfo{person}{Nathan Srebro}.} \bibinfo{year}{2005}\natexlab{a}.
\newblock \showarticletitle{Fast maximum margin matrix factorization for
  collaborative prediction}. In \bibinfo{booktitle}{\emph{Proceedings of the
  22nd international conference on Machine learning}}. ACM,
  \bibinfo{pages}{713--719}.
\newblock


\bibitem[\protect\citeauthoryear{Rennie and Srebro}{Rennie and Srebro}{2005b}]%
        {rennie2005loss}
\bibfield{author}{\bibinfo{person}{Jason~DM Rennie} {and}
  \bibinfo{person}{Nathan Srebro}.} \bibinfo{year}{2005}\natexlab{b}.
\newblock \showarticletitle{Loss functions for preference levels: Regression
  with discrete ordered labels}. In \bibinfo{booktitle}{\emph{Proceedings of
  the IJCAI multidisciplinary workshop on advances in preference handling}}.
  Kluwer Norwell, MA, \bibinfo{pages}{180--186}.
\newblock


\bibitem[\protect\citeauthoryear{Stekhoven}{Stekhoven}{2011}]%
        {stekhoven2011using}
\bibfield{author}{\bibinfo{person}{Daniel~J Stekhoven}.}
  \bibinfo{year}{2011}\natexlab{}.
\newblock \showarticletitle{Using the missForest package}.
\newblock \bibinfo{journal}{\emph{R package}} (\bibinfo{year}{2011}),
  \bibinfo{pages}{1--11}.
\newblock


\bibitem[\protect\citeauthoryear{Stekhoven and B{\"u}hlmann}{Stekhoven and
  B{\"u}hlmann}{2011}]%
        {stekhoven2011missforest}
\bibfield{author}{\bibinfo{person}{Daniel~J Stekhoven} {and}
  \bibinfo{person}{Peter B{\"u}hlmann}.} \bibinfo{year}{2011}\natexlab{}.
\newblock \showarticletitle{MissForest—non-parametric missing value
  imputation for mixed-type data}.
\newblock \bibinfo{journal}{\emph{Bioinformatics}} \bibinfo{volume}{28},
  \bibinfo{number}{1} (\bibinfo{year}{2011}), \bibinfo{pages}{112--118}.
\newblock


\bibitem[\protect\citeauthoryear{Tsukahara}{Tsukahara}{2005}]%
        {tsukahara2005semiparametric}
\bibfield{author}{\bibinfo{person}{Hideatsu Tsukahara}.}
  \bibinfo{year}{2005}\natexlab{}.
\newblock \showarticletitle{Semiparametric estimation in copula models}.
\newblock \bibinfo{journal}{\emph{Canadian Journal of Statistics}}
  \bibinfo{volume}{33}, \bibinfo{number}{3} (\bibinfo{year}{2005}),
  \bibinfo{pages}{357--375}.
\newblock


\bibitem[\protect\citeauthoryear{Turnbull, Barrington, Torres, and
  Lanckriet}{Turnbull et~al\mbox{.}}{2007}]%
        {turnbull2007towards}
\bibfield{author}{\bibinfo{person}{Douglas Turnbull}, \bibinfo{person}{Luke
  Barrington}, \bibinfo{person}{David Torres}, {and} \bibinfo{person}{Gert
  Lanckriet}.} \bibinfo{year}{2007}\natexlab{}.
\newblock \showarticletitle{Towards musical query-by-semantic-description using
  the cal500 data set}. In \bibinfo{booktitle}{\emph{Proceedings of the 30th
  annual international ACM SIGIR conference on Research and development in
  information retrieval}}. \bibinfo{pages}{439--446}.
\newblock


\bibitem[\protect\citeauthoryear{Udell, Horn, Zadeh, Boyd, et~al\mbox{.}}{Udell
  et~al\mbox{.}}{2016}]%
        {udell2016generalized}
\bibfield{author}{\bibinfo{person}{Madeleine Udell}, \bibinfo{person}{Corinne
  Horn}, \bibinfo{person}{Reza Zadeh}, \bibinfo{person}{Stephen Boyd},
  {et~al\mbox{.}}} \bibinfo{year}{2016}\natexlab{}.
\newblock \showarticletitle{Generalized low rank models}.
\newblock \bibinfo{journal}{\emph{Foundations and Trends{\textregistered} in
  Machine Learning}} \bibinfo{volume}{9}, \bibinfo{number}{1}
  (\bibinfo{year}{2016}), \bibinfo{pages}{1--118}.
\newblock


\bibitem[\protect\citeauthoryear{Udell and Townsend}{Udell and
  Townsend}{2019}]%
        {udell2019why}
\bibfield{author}{\bibinfo{person}{Madeleine Udell} {and} \bibinfo{person}{Alex
  Townsend}.} \bibinfo{year}{2019}\natexlab{}.
\newblock \showarticletitle{Why are Big Data Matrices Approximately Low Rank?}
\newblock \bibinfo{journal}{\emph{SIAM Journal on Mathematics of Data Science
  (SIMODS)}} \bibinfo{volume}{1}, \bibinfo{number}{1} (\bibinfo{year}{2019}),
  \bibinfo{pages}{144--160}.
\newblock
\urldef\tempurl%
\url{https://epubs.siam.org/doi/pdf/10.1137/18M1183480}
\showURL{%
\tempurl}


\bibitem[\protect\citeauthoryear{Vaart and Wellner}{Vaart and Wellner}{1996}]%
        {vaart1996weak}
\bibfield{author}{\bibinfo{person}{Aad~W Vaart} {and} \bibinfo{person}{Jon~A
  Wellner}.} \bibinfo{year}{1996}\natexlab{}.
\newblock \bibinfo{booktitle}{\emph{Weak convergence and empirical processes:
  with applications to statistics}}.
\newblock \bibinfo{publisher}{Springer}.
\newblock


\bibitem[\protect\citeauthoryear{Van~Buuren and Oudshoorn}{Van~Buuren and
  Oudshoorn}{1999}]%
        {van1999flexible}
\bibfield{author}{\bibinfo{person}{Stef Van~Buuren} {and}
  \bibinfo{person}{Karin Oudshoorn}.} \bibinfo{year}{1999}\natexlab{}.
\newblock \bibinfo{booktitle}{\emph{Flexible multivariate imputation by MICE}}.
\newblock \bibinfo{publisher}{Leiden: TNO}.
\newblock


\bibitem[\protect\citeauthoryear{Wang, Wang, Yang, and Wang}{Wang
  et~al\mbox{.}}{2014}]%
        {wang2014towards}
\bibfield{author}{\bibinfo{person}{Shuo-Yang Wang}, \bibinfo{person}{Ju-Chiang
  Wang}, \bibinfo{person}{Yi-Hsuan Yang}, {and} \bibinfo{person}{Hsin-Min
  Wang}.} \bibinfo{year}{2014}\natexlab{}.
\newblock \showarticletitle{Towards time-varying music auto-tagging based on
  CAL500 expansion}. In \bibinfo{booktitle}{\emph{2014 IEEE International
  Conference on Multimedia and Expo (ICME)}}. IEEE, \bibinfo{pages}{1--6}.
\newblock


\end{thebibliography}

\clearpage
\appendix
\section{Computational Details}
\subsection{Details for Section \ref{sec:comp_ordinal}}
Denote the observation $\{\xobs,\Sigma\}$ i.e. $\{\bz_{\indexC}=\bigf_{\indexC}^{-1}(\bx_{\indexC}), \bz_{\indexD}\in\bigf_{\indexD}^{-1}(\bx_{\indexD}), \Sigma\}$ as $\{\bast\}$. Since the task is to compute the marginal mean and variance of a multivariate truncated normal, we suppose $\indMis=\emptyset$ here without loss of generality. For each $j\in\indexD$, we use the law of total expectation by conditioning on $\zdj$ first. 
Given $\{\bast, \zdj\}$, $z_j$ is univariate normal with mean $\tilde\mu_j=\SigmaJO\SigmaNJJ^{-1}\bz_{-j}$ and variance $\tilde\sigma_j^2=1-\SigmaJO\SigmaNJJ^{-1}\SigmaOJ$, truncated to the region $f_j^{-1}(x_j)$,
where the index $-j$ means all dimensions but $j$, i.e., $[p] \setminus {j}$. 
The region $f_j^{-1}(x_j)$ is an interval:  $f_j^{-1}(x_j)=(a_j,b_j]$. Here are three cases: (1) $a_j,b_j \in \mathbb{R}$; (2) $a_j\in \mathbb{R}, b_j=\infty$; (3) $a_j=-\infty, b_j\in \mathbb{R}$. 
The computation for all cases are similar. We take the first case as an example. 
First we introduce a lemma describing the first and second moments of a truncated univariate normal.
\begin{lemma}
	Suppose a univariate random variable $z\sim \mathcal{N}(\mu,\sigma^2)$. For constants $a<b$, let $\alpha=(a-\mu)/\sigma$ and $\beta=(b-\mu)/\sigma$. Then the mean and variance of $z$ truncated to the interval $(a,b]$ are:
	\begin{equation*}
		\Erm(z|a<z\leq b) = \mu +\frac{\phi(\alpha)-\phi(\beta)}{\Phi(\beta)-\Phi(\alpha)} \cdot \sigma
		\label{Eq:mean_tmean}
	\end{equation*} 
		\begin{equation*}
	\Varrm (z|a<z\leq b) =\left( 1+ \frac{\alpha\phi(\alpha)-\beta\phi(\beta)}{\Phi(\beta)-\Phi(\alpha)}- \left(\frac{\phi(\alpha)-\phi(\beta)}{\Phi(\beta)-\Phi(\alpha)} \right)^2  \right)\sigma^2.
		\label{Eq:mean_tvar}
	\end{equation*} 
	\label{lemma:truncated}
\end{lemma}
Plugging $\mu =\tilde\mu_j, \sigma^2=\tilde \sigma_j^2$ and $(a,b]=f_j^{-1}(x_j)$ 
into the above mean and variance formulas,
we obtain the expression of $g_j(\bz_{\indexD-j};x_j,\Sigma)$ defined in Section \ref{sec:comp_ordinal},
and the univariate truncated normal variance $\Varrm[z_j|\bz_{\indexD-j}, \xobs,\Sigma]=:h_j(\bz_{\indexD-j};x_j,\Sigma)$, a nonlinear function  $\mathbb{R}^{|\indexD|-1}\rightarrow  \mathbb{R}$,
parameterized by $x_j$ and $\Sigma$.
Write down the formula for marginal variance conditional on observation:
\begin{align*}
	\Varrm[z_j|\bast]&= {\Erm\left[  \Varrm[z_j|\zdj, \bast] \big| \bast\right]} + 	\Varrm\left[  \Erm[z_j|\zdj, \bast] \big| \bast\right] \\
	&=\Erm\left[  h_j(\zdj;x_j,\Sigma)\big|\bast \right]
	+ \Varrm\left[  g_j(\zdj;x_j,\Sigma)\big|\bast \right]
\end{align*}
We approximate the first term as $h_j(\Erm[\zdj| \bast];x_j,\Sigma)$. As for the second term, \citet{guo2015graphical} approximated it as $\Varrm[\tilde \mu_j|\bast]$ based on $\Erm\left[  g_j^2(\zdj;x_j,\Sigma)\big|\bast \right] \approx g_j^2(\Erm[\zdj| \bast];x_j,\Sigma)$. However, we found in practice simply dropping the second term performs better.

In summary, given an estimate $\hat \bz_\indexD^{(t)} \approx \Erm[\bz_\indexD|\xobs,\Sigma^{(t)}]$ and $\Sigma^{(t+1)}$, for $j\in \indexD$, we update $\Erm[z_j|\xobs,\Sigma^{(t+1)}]\approx g_j(\hat \bz_{\indexD-j}^{(t)}; x_j,\Sigma^{(t+1)})$ and \\
 $\Varrm[z_j|\xobs,\Sigma^{(t+1)}]\approx h_j(\hat \bz_{\indexD-j}^{(t)}; x_j,\Sigma^{(t+1)})$. In other words, we update the conditional mean and variance of $z_j$ as the univariate truncated normal mean and variance with all other observed ordinal dimensions equal to their mean from last iteration, i.e. $\bz_{\indexD-j}=\hat \bz_{\indexD-j}^{(t)}$.

\subsection{Details for Section \ref{sec:comp_em}}
Given $\Erm[\bz_{\indexO}| \bast], \Erm[\bz_{\indMis}| \bast]$ and $\Covrm[\bz_{\indexO}| \bast]$,
it suffices to compute $\Erm[\bz_\indMis\bz^\intercal_{\indObs} |\bast]$ and $\Erm[\bz_\indMis\bz^\intercal_{\indMis} |\bast]$ for $\Covrm[\bz_\indMis,\bz_{\indObs} | \bast]$ and $ \Covrm[\bz_\indMis| \bast]$. Using the law of total expectation, we have:
\begin{align*}
  &\Erm[\bz_\indMis\bz^\intercal_{\indObs} | \bast] = \Erm\left[\Erm[\bz_\indMis\bz^\intercal_{\indObs} |\bz_{\indObs}, \bast] \Big|\bast \right] = \Erm\left[\Erm[\bz_\indMis |\bz_{\indObs}, \bast]\cdot \bz^\intercal_{\indObs} \Big|\bast \right]\nonumber\\
  =&  \Erm\left[ \SigmaMO\SigmaOO^{-1}\bz_{\indObs}\cdot \bz^\intercal_{\indObs} \Big|\bast \right] = \SigmaMO\SigmaOO^{-1} \Erm[\bz_{\indObs}\bz^\intercal_{\indObs} | \bast].
\end{align*}
\begin{align*}
&\Erm[\bz_\indMis\bz^\intercal_{\indMis} | \bast] = \Erm\left[ \Erm[\bz_\indMis\bz^\intercal_{\indMis} |\bz_{\indObs}, \bast] \Big|\bast \right] \nonumber \\
=&  \Erm\left[ \Covrm[\bz_\indMis|\bz_{\indObs}, \bast] \Big|\bast \right] + \Erm\left[ \Erm[\bz_\indMis|\bz_{\indObs}, \bast]\cdot \Erm[\bz^\intercal_\indMis|\bz_{\indObs}, \bast] \Big|\bast \right]\nonumber\\
=& \SigmaMM-\SigmaMO\SigmaOO^{-1}\SigmaOM + \Sigma_{\indMis,\indObs}\Sigma_{\indObs,\indObs}^{-1} \Erm[\bz_\indObs | \bast] \Erm[\bz^\intercal_{\indObs} | \bast] \SigmaOO^{-1}\SigmaOM.
\end{align*}

\section{Supplement for Experiments}
\subsection{Implementation Details}
For \texttt{softImpute},
we first center the rows and columns,
then select the penalization parameter in the path from $45$ (rank $12$) to $6$ (rank $207$) with $50$ points.
For \texttt{GLRM},
we use quadratic regularization on $X$ factor and ordinal regularization on $Y$ factor. 
The model is fitted with SVD initialization and offest term.
After a small grid search, we select the quadratic regularization parameter as $n_\textup{obs}\times 1.2\times 10^{-4}$ where $n_\textup{obs}$ is the number of observed entries. 
Then the rank is selected through an exhaustive search.
For \texttt{xPCA} and \texttt{imputeFAMD}, the rank is selected through an exhaustive search.

\subsection{Results of \texttt{sbgcop} on Real Datasets}
For GSS data, \texttt{Copula-EM} takes 24s, while \texttt{sbgcop} with $1000$ iterations takes 87s, with imputation error: $\texttt{CALSS}$, $0.992(0.13)$; $\texttt{LIFE}$, $0.924(0.7)$; $\texttt{HEALTH}$, $1.132(0.15)$; $\texttt{HAPPY}$, $1.231(0.11)$; $\texttt{INCOME}$, $0.931(0.03)$. 

For movielens data, \texttt{Copula-EM} takes 9 mins, while \texttt{sbgcop} with $200$ iterations takes 33 mins, with imputation error: MAE, $0.752(0.004)$; RMSE, $1.030(0.005)$. 

For CAL500exp data, \texttt{Copula-EM} takes 80s, while \texttt{sbgcop} with $500$ iterations takes 290s, with imputation error: $1.301(0.019)$ for $40\%$ missing ratio; $1.328(0.015)$ for $50\%$ missing ratio; $1.379(0.016)$ for $60\%$ missing ratio.

For four small datasets used in Section \ref{sec:moredata}, the time \texttt{sbgcop} with $1000$ iterations takes is 2 times to 9 times (varying over datasets) of the time \texttt{Copula-EM} takes. 
The corresponding imputation error is: 
ESL label $0.466(0.04)$, feature $0.649(0.02)$; 
LEV label $0.849(0.03)$, feature $0.936(0.01)$; 
GBSG ordinal $0.992(0.03)$, continuous $0.953(0.02)$;
TIPS ordinal $0.984(0.06)$, continuous $0.768(0.05)$.
	
	\subsection{Datasets Description for Section \ref{sec:moredata}}
	\begin{description}
  \item[\textit{ESL}] This dataset contains profiles of applicants for certain
	jobs. The recruiting company, based upon psychometric
	 test results and interviews with the candidates, determined the values of the
	input attributes. The output is an overall score corresponding to the
	degree of fitness of the candidate.
  \item[\textit{LEV}] This dataset contains lecturer evaluations. 
  Students evaluate their lecturers according to four attributes such as 
	oral skills and contribution to their professional/general knowledge. 
	The output is an overall score of the lecturer’s performance. 
	\item[\textit{GBSG}] This dataset contains the information of women with breast 
	cancer concerning the status of the tumours and the hormonal system of the patient. 
	\item[\textit{TIPS}]This dataset concerns the tips given to a waiter in a restaurant collected from customers. Recording variables contains the price of the meal, the tip amount and the conditions of the restaurant meal (number of guests, time of data, etc.). 
\end{description}

\clearpage
\section{Proof of Lemmas}
\paragraph{Proof of Lemma 1}
\begin{proof}
For any $j\in[p]$, $x_j\equald f_j(z_j)$ if and only if (iff) $x_j$ and $f_j(z_j)$ have the same CDF. For each $j\in[p]$, since $f_j^{-1}$ exists for any strictly monotone $f_j$, we can calculate the CDF of $f_j(z_j)$:
\begin{align*}
	F_{f_j(z_j)}(t)=\bP(f_j(z_j)\leq t)=\bP(z_j\leq f_j^{-1}(t))=\Phi(f_j^{-1}(t)).
\end{align*}
Then $x_j\equald f_j(z_j)$ iff $\Phi\circ f_j^{-1}=F_j$, equivalently, $f_j=F^{-1}_j\circ \Phi$.
\end{proof}

\paragraph{Proof of Lemma 2}
\begin{proof}
It suffices to show for monotone function $f$, $x\equald f(z)$ iff $f(z)=\cutoff(z; \bS)$
with $\bS=\{s_l = F_z^{-1}\left( \sum_{t=1}^lp_t\right): l \in [k-1]\}$. Notice $x\equald f(z)$ iff the range of $f(z)$ is $[k]$ and $p_l=\bP(f(z)=l)$ for any $l\in[k]$. When $f(z)=\cutoff(z; \bS)$, further define $s_k=\infty$ and $s_0=-\infty$. Since $z$ is continuous with CDF $F_z$,
it suffices to show:
\[
\bP(f(z)=l) = \bP(s_{l-1}<z\leq s_l)=F_z(s_l) - F_z(s_{l-1})=p_l, \mbox{ for }l\in [k]
\]

When $x\equald f(z)$, $f(z)$ has range $[k]$. For $l\in[k]$, define $A_l=\{z:f(z)=l\}, s_l=\sup\limits_{z\in A_l}z$ and $s_0=\inf\limits_{z\in A_1}z$. Since $\bP(f(z)=l)=p_l>0$, we have $\inf\limits_{z\in A_l}z<s_l$. Since $f$ is monotone, we have $s_{l-1}\leq \inf\limits_{z\in A_l}z$. Claim $s_{l-1}= \inf\limits_{z\in A_l}z$. If not, there exists $s_{l-1}<z^*<\inf\limits_{z\in A_l}z$ satisfying $(l-1)\leq f(z^*)\leq l$. Since $f(z)$ has range $[k]$, $f(z^*)$ can only be $l$ or $l-1$. Equivalently $z^*\in A_l$ or  $z^*\in A_{l-1}$, which contradicts $s_{l-1}<z^*<\inf\limits_{z\in A_l}z$. Thus $s_{l-1}=\inf\limits_{z\in A_l}z$, $f(z)=1+\sum_{l=1}^{k-1}\mathds{1}(z > s_l)$,
\[
p_l=\bP(f(z)=l) = \bP(z\in A_l) = \bP(s_{l-1}\leq z\leq s_l) = F_z(s_l)-F_z(s_{l-1}),
\]
Thus we have
$F_z(s_l) = \sum_{t=1}^lp_t \Rightarrow s_l=F_z^{-1}(\sum_{t=1}^lp_t)$.

\end{proof}

\paragraph{Proof of Lemma 3}
Before we prove Lemma 3, we introduce the Dvoretzky-Kiefer-Wolfowitz inequality  \cite{dvoretzky1956asymptotic}, also introduced in \cite{kosorok2008introduction}.
\begin{inequalityContinuous}
		For any i.i.d. sample $x^1,\ldots,x^n$ with distribution $F$, then when $\epsilon>0$, 
	\[
	\bP\left(\sup_{t\in\mathbb{R}}|\mathbb{F}_n(t)-F(t)|\geq \epsilon \right) \leq 2e^{-2n\epsilon^2}, \mbox{ where } \mathbb{F}_n(t)=\frac{\sum_{i=1}^n1\{x^i\leq t\}}{n}
	\]
	\label{lemma:continuous}
\end{inequalityContinuous}
\begin{proof}
Applying the Dvoretzky-Kiefer-Wolfowitz inequality, for any $\epsilon >0$,
$\Pr(\sup_{t\in\mathbb{R}}|\mathbb{F}_n(t)-F(t)|< \epsilon)\geq 1--2e^{-2n\epsilon^2}$. 

Take $\epsilon > n^{-1}$, $\sup_{t\in\mathbb{R}}\left|\frac{n}{n+1}\mathbb{F}_n(t)-F(t)\right|  <2\epsilon$.
Further let  $\epsilon < K_1\triangleq \min{\{\frac{F(m)}{4}, \frac{1-F(M)}{4}}\}$, 
we have $\frac{n}{n+1}\mathbb{F}_n(t)\in [\frac{F(m)}{2}, \frac{1+F(M)}{2}]$ 
for $t\in [m,M]$.
Then,
\begin{align*}
&\sup_{t\in [m,M]}	\left|\hat f^{-1}(t)-f^{-1}(t) \right| = \sup_{t\in [m,M]}	\left|\Phi^{-1}\left( \frac{n}{n+1}\mathbb{F}_n(t)\right) -\Phi^{-1}(F(t)) \right| \\
&\leq \sup_{r\in [\frac{F(m)}{2}, \frac{1+F(M)}{2}]}\left|\left( \Phi^{-1}(r) \right)'\right|\cdot   \sup_{t\in [m,M]}\left|\frac{n}{n+1}\mathbb{F}_n(t)-F(t)\right| \\
&< 2\epsilon \cdot  \sup_{r\in [\frac{F(m)}{2}, \frac{1+F(M)}{2}]}\left|\left( \Phi^{-1}(r) \right)'\right|
\end{align*}
Since $ \left( \Phi^{-1}(r) \right)'=\frac{1}{\phi(\Phi^{-1}(r))}$, we get $ \sup_{r\in [\frac{F(m)}{2}, \frac{1+F(M)}{2}]}\left|\left( \Phi^{-1}(r) \right)'\right|=K_2\triangleq1/\min\left\{\phi\left(\Phi^{-1}(\frac{F(m)}{2}) \right), \phi\left(\Phi^{-1}(\frac{F(M)+1}{2}) \right) \right \}$. Adjusting the constants, for $2K_2n^{-1}<\epsilon<2K_1K_2$, we have
\[
\bP\left(	\sup_{t\in [m,M]}	\left|\hat f^{-1}(t)-f^{-1}(t) \right|  >\epsilon\right) \leq 2\exp\left\{-\frac{n\epsilon^2}{2K_2^2}\right\}.
\]
\end{proof}

\paragraph{Proof of Lemma 4}
Before we prove Lemma 4, we introduce the Bretagnolle-Huber-Carol inequality introduced in \cite[]{vaart1996weak}.
\begin{inequalityOrdinal}
		If the random vector $(N_1,\ldots,N_k)$ is multinomially distributed with parameters $n$ and $(p_1,\ldots,p_k)$, then
	\[
	\bP\left(\sum_{i=1}^k|N_i/n-p_i|\geq \epsilon \right) \leq 2^ke^{-\frac{1}{2}n\epsilon^2}, \qquad \epsilon>0.
	\]
	\label{lemma:ordinal}
\end{inequalityOrdinal}
\begin{proof}

According to Lemma 2, the cutoff function $f(z)=\cutoff(z;\bS)$ is unique and $\bS=\{s_l: s_l = \Phi^{-1}(\sum_{t=1}^lp_t), l\in [k-1]\}$. Define $s^*_l=\Phi^{-1}\left(\frac{\sum_{i=1}^n\mathds{1}(x^i\leq l)}{n} \right)$ for $l\in [k-1]$, $s_0^*=-\infty, s_k^*=\infty$, and $\deltastar = \Phi(s^*_l)-\Phi(s^*_{l-1})=\sum_{i=1}^n\mathds{1}(x^i= l)/n$. Notice $(n\Delta^*_1,\ldots,n\Delta^*_k)$ is multinomially distributed with parameters $n$ and $(p_1,\ldots,p_k)$, applying the Bretagnolle-Huber-Carol inequality, for any $\epsilon>0$, with probability at least $1-2^ke^{-\frac{1}{2}n\epsilon^2}$,
$\sum_{l=1}^k|\deltastar-p_l|< \epsilon$. First for each $l\in [k]$, $|\Phi(s_l^*) - 	\Phi(s_l)|\leq \sum_{t=1}^k|\Delta^*_t- p_t|<\epsilon$.
Take $\epsilon > n^{-1}$, we have
\begin{align*}
	\left| \Phi(s_l^*)\cdot \frac{n}{n+1} - 	\Phi(s_l)\right|\leq 	|\Phi(s_l^*) - 	\Phi(s_l)| + 	\frac{ \Phi(s_l^*)}{n+1}< 2\epsilon
\end{align*}
$$
\Phi(s_l)-2\epsilon <  \Phi(s_l^*)\cdot \frac{n}{n+1} = \frac{\sum_{i=1}^n\mathds{1}(x^i\leq l)}{n+1} < \Phi(s_l)+2\epsilon
$$
When $l\in[k-1]$, we have $p_1\leq \Phi(s_l)\leq \sum_{t=1}^{k-1}p_t$. Further let $\epsilon < K_1\triangleq \min{\{\frac{p_1}{4}, \frac{p_k}{4}}\}$, we have $\frac{p_1}{2} \leq \Phi(s_l^*)\cdot \frac{n}{n+1} \leq 1-\frac{p_k}{2}$. Thus:
\begin{align*}
	||\hat \bS - \bS||_1&=\sum_{l=1}^{k-1}\left|\hat s_l - s_l \right| = 		\sum_{l=1}^{k-1}\left|\Phi^{-1}\left(\frac{\sum_{i=1}^n\mathds{1}(x^i\leq l)}{n+1} \right)  - \Phi^{-1}(\Phi(s_l)) \right|\\
	&\leq \sup_{r\in [\frac{p_1}{2}, 1-\frac{p_k}{2}]}\left|\left( \Phi^{-1}(r) \right)'\right|\cdot 	\sum_{l=1}^{k-1}\left|\frac{\sum_{i=1}^n\mathds{1}(x^i\leq l)}{n+1}  -\Phi(s_l) \right| \\
	&\leq \frac{1}{\min\left\{\phi\left(\Phi^{-1}(\frac{p_1}{2}) \right), \phi\left(\Phi^{-1}(1-\frac{p_k}{2}) \right) \right \}} \cdot 2(k-1)\epsilon
\end{align*}
Let  $K_2=1/\min\left\{\phi\left(\Phi^{-1}(\frac{p_1}{2}) \right), \phi\left(\Phi^{-1}(1-\frac{p_k}{2}) \right) \right \} $. Adjusting the constants, for $2(k-1)K_2n^{-1}<\epsilon<2(k-1)K_1K_2$,  we have
\[
\bP\left(	||\hat \bS - \bS||_1|  >\epsilon\right) \leq 2\exp\left\{-\frac{1}{8K_2^2}\cdot \frac{n\epsilon^2}{(k-1)^2}\right\}.
\]
\end{proof}

\end{document}